%% file: VandenBerk.tex
\documentclass[preprint]{aastex}

\newcommand{\cii}{C\,{\sc ii}\ }
\newcommand{\ciii}{C\,{\sc iii}]\ }
\newcommand{\civ}{C\,{\sc iv}\ }

\newcommand{\mgii}{Mg\,{\sc ii}\ }
\newcommand{\feii}{Fe\,{\sc ii}\ }
\newcommand{\feiii}{Fe\,{\sc iii}\ }
\newcommand{\caii}{Ca\,{\sc ii}\ }
\newcommand{\lya}{Ly\,$\alpha$\ }
\newcommand{\halpha}{H\,$\alpha$\ }
\newcommand{\hbeta}{H\,$\beta$\ }

\newcommand{\oiii}{[O\,{\sc iii}]\ }

\newcommand{\nv}{N\,{\sc v}\ }

\begin{document}

\title{Composite Quasar Spectra From the
  Sloan Digital Sky Survey\altaffilmark{1}}

\author{
Daniel E. Vanden Berk\altaffilmark{2},
Gordon T. Richards\altaffilmark{3},
Amanda Bauer\altaffilmark{4},
Michael A. Strauss\altaffilmark{5},
Donald P. Schneider\altaffilmark{3},
Timothy M. Heckman\altaffilmark{6},
Donald G. York\altaffilmark{7,8},
Patrick B. Hall\altaffilmark{5},
Xiaohui Fan\altaffilmark{5,9},
G. R. Knapp\altaffilmark{5},
Scott F. Anderson\altaffilmark{10},
James Annis\altaffilmark{2},
Neta A. Bahcall\altaffilmark{5},
Mariangela Bernardi\altaffilmark{7},
John W. Briggs\altaffilmark{7},
J. Brinkmann\altaffilmark{11},
Robert Brunner\altaffilmark{12},
Scott Burles\altaffilmark{2},
Larry Carey\altaffilmark{10},
Francisco J. Castander\altaffilmark{7,13},
A. J. Connolly\altaffilmark{14},
J. H. Crocker\altaffilmark{6},
Istv\'an Csabai\altaffilmark{6,15},
Mamoru Doi\altaffilmark{16},
Douglas Finkbeiner\altaffilmark{17},
Scott Friedman\altaffilmark{6},
Joshua A. Frieman\altaffilmark{2,7},
Masataka Fukugita\altaffilmark{18},
James E. Gunn\altaffilmark{5},
G. S. Hennessy\altaffilmark{19},
\v{Z}eljko Ivezi\'{c}\altaffilmark{5},
Stephen Kent\altaffilmark{2,7},
Peter Z. Kunszt\altaffilmark{6},
D.Q. Lamb\altaffilmark{7},
R. French Leger\altaffilmark{10},
Daniel C. Long\altaffilmark{11},
Jon Loveday\altaffilmark{20},
Robert H. Lupton\altaffilmark{5},
Avery Meiksin\altaffilmark{21},
Aronne Merelli\altaffilmark{12,22},
Jeffrey A. Munn\altaffilmark{23},
Heidi Jo Newberg\altaffilmark{24},
Matt Newcomb\altaffilmark{22},
R. C. Nichol\altaffilmark{22},
Russell Owen\altaffilmark{10},
Jeffrey R. Pier\altaffilmark{23},
Adrian Pope\altaffilmark{6,22},
Constance M. Rockosi\altaffilmark{7},
David J. Schlegel\altaffilmark{5},
Walter A. Siegmund\altaffilmark{10},
Stephen Smee\altaffilmark{6,25},
Yehuda Snir\altaffilmark{22},
Chris Stoughton\altaffilmark{2},
Christopher Stubbs\altaffilmark{10},
Mark SubbaRao\altaffilmark{7},
Alexander S. Szalay\altaffilmark{6},
Gyula P. Szokoly\altaffilmark{6},
Christy Tremonti\altaffilmark{6},
Alan Uomoto\altaffilmark{6},
Patrick Waddell\altaffilmark{10},
Brian Yanny\altaffilmark{2},
Wei Zheng\altaffilmark{6}
}

\altaffiltext{1}{Based on observations obtained with the Sloan Digital
Sky Survey, which is owned and operated by the Astrophysical Research
Consortium.}
\altaffiltext{2}{Fermi National Accelerator Laboratory, P.O. Box 500, Batavia, IL 60510}
\altaffiltext{3}{Department of Astronomy and Astrophysics, The Pennsylvania State University, University Park, PA 16802}
\altaffiltext{4}{Department of Physics, University of Cincinnati, 400 Physics Bldg., Cincinnati, OH 45221}
\altaffiltext{5}{Princeton University Observatory, Princeton, NJ 08544}
\altaffiltext{6}{Department of Physics and Astronomy, The Johns Hopkins University, 3701 San Martin Drive, Baltimore, MD 21218}
\altaffiltext{7}{The University of Chicago, Department of Astronomy and Astrophysics, 5640 S. Ellis Ave., Chicago, IL 60637}
\altaffiltext{8}{The University of Chicago, Enrico Fermi Institute, 5640 S. Ellis Ave., Chicago, IL 60637}
\altaffiltext{9}{Institute for Advanced Study, Olden Lane, Princeton, NJ 08540}
\altaffiltext{10}{University of Washington, Department of Astronomy, Box 351580, Seattle, WA 98195}
\altaffiltext{11}{Apache Point Observatory, P.O. Box 59, Sunspot, NM 88349-0059}
\altaffiltext{12}{Department of Astronomy, California Institute of Technology, Pasadena, CA 91125}
\altaffiltext{13}{Observatoire Midi Pyrenees, 14 ave Edouard Belin, Toulouse, F-31400, France}
\altaffiltext{14}{Department of Physics and Astronomy, University of Pittsburgh, Pittsburgh, PA 15260}
\altaffiltext{15}{Department of Physics of Complex Systems, E\"otv\"os University, P\'azm\'any P\'eter s\'et\'any 1}
\altaffiltext{16}{Department of Astronomy and Research Center for the Early Universe, School of Science, University of Tokyo, Hongo, Bunkyo, Tokyo, 113-0033, Japan}
\altaffiltext{17}{University of California at Berkeley, Departments of Physics and Astronomy, 601 Campbell Hall, Berkeley, CA 94720}
\altaffiltext{18}{University of Tokyo, Institute for Cosmic Ray Reserach, Kashiwa, 2778582, Japan}
\altaffiltext{19}{U.S. Naval Observatory, 3450 Massachusetts Ave., NW, Washington, DC  20392-5420}
\altaffiltext{20}{Astronomy Centre, University of Sussex, Falmer, Brighton BN1 9QJ, United Kingdom}
\altaffiltext{21}{Royal Observatory, Edinburgh, EH9 3HJ, United Kingdom}
\altaffiltext{22}{Dept. of Physics, Carnegie Mellon University, 5000 Forbes Ave., Pittsburgh, PA-15232}
\altaffiltext{23}{U.S. Naval Observatory, Flagstaff Station, P.O. Box 1149, Flagstaff, AZ  86002-1149}
\altaffiltext{24}{Physics Department, Rensselaer Polytechnic Institute, SC1C25, Troy, NY 12180}
\altaffiltext{25}{Department of Astronomy, University of Maryland, College Park, MD 20742-2421}

\begin{abstract}
We have created a variety of composite quasar spectra using a
homogeneous data set of over $2200$ spectra from the Sloan Digital Sky
Survey (SDSS).  The quasar sample spans a redshift range of $0.044 \leq
z \leq 4.789$, and an absolute $r'$ magnitude range of $-18.0$ to
$-26.5$.  The input spectra cover an observed wavelength range of $3800
- 9200${\AA} at a resolution of $1800$. The median composite covers a
rest wavelength range from $800 - 8555$\,{\AA}, and reaches a peak
signal-to-noise ratio of over 300 per 1\,{\AA} resolution element in
the rest frame.  We have identified over 80 emission line features in
the spectrum.  Emission line shifts relative to nominal laboratory
wavelengths are seen for many of the ionic species.  Peak shifts of the
broad permitted and semi-forbidden lines are strongly correlated with
ionization energy, as previously suggested, but we find that the narrow
forbidden lines are also shifted by amounts which are strongly
correlated with ionization energy.  The magnitude of the forbidden line
shifts is $\la 100$\,km\,s$^{-1}$, compared to shifts of up to
$550$\,km\,s$^{-1}$ for some of the permitted and semi-forbidden
lines.  At wavelengths longer than the \lya emission, the continuum of
the geometric mean composite is well-fit by two power-laws, with a
break at $\approx 5000$\,{\AA}.  The frequency power law index,
$\alpha_\nu$, is $-0.44$ from $\approx 1300 - 5000$\,{\AA}, and $-2.45$
redward of $\approx 5000$\,{\AA}.  The abrupt change in slope can be
accounted for partly by host galaxy contamination at low redshift.
Stellar absorption lines, including higher-order Balmer lines, seen in
the composites suggest that young or intermediate age stars make a
significant contribution to the light of the host galaxies.  Most of
the spectrum is populated by blended emission lines, especially in the
range $1500 - 3500${\AA}, which can make the estimation of quasar
continua highly uncertain unless large ranges in wavelength are
observed.  An electronic table of the median quasar template is
available.  \end{abstract}


\section{Introduction \label{intro}}
Most quasar spectra from ultraviolet to optical wavelengths can be
characterized by a featureless continuum and a series of mostly broad
emission line features; compared with galaxies or stars, these spectra
are remarkably similar from one quasar to another.  The first three
principal components spectra account for about 75\% of the intrinsic
quasar variance \citep{francis92}.  Subtle global spectral properties
can be studied by combining large numbers of quasar spectra into
composites.  The most detailed composites \citep{francis91,
zheng97, brotherton00} use hundreds of moderate resolution spectra, and
typically cover a few thousand {\AA} in the quasar rest frame.  These high
signal-to-noise ratio (S/N) spectra reveal variations from a single power law
in the general continuum shape, and weak emission features that are
rarely detectable in individual quasar spectra.

The Sloan Digital Sky Survey \citep[][SDSS]{york00} already
contains spectra for over 2500 quasars as of June 2000, and by survey
end will include on the order of $10^5$ quasar spectra.  The
identification and basic measurement of this sample will be done using
an automated pipeline, part of which uses templates for object
classification and redshift determination.  As one of the first uses of
the initial set of spectra, we have created a composite quasar spectrum
for use as a template.  The large number of spectra, their wavelength
coverage, relatively high resolution, and high signal-to-noise ratio,
make the current SDSS sample ideal for the creation of composite quasar
spectra.  The resulting composite spectrum covers a vacuum rest
wavelength range of $800 - 8555$\,{\AA}. The peak S/N per $1$\,{\AA}
resolution element is over $300$ near $2800${\AA} -- several times
higher than the previous best ultraviolet/optical composites
\citep[e.g.][]{francis91, zheng97, brotherton00}.

In addition to serving as a cross-correlation template, the composite
is useful for the precise measurement of emission line shifts relative
to nominal laboratory wavelengths, calculation of quasar colors for
improved candidate selection and photometric redshift estimates, the
calculation of K-corrections used in evaluating the quasar luminosity
function, and for the estimation of the backlighting flux density
continuum for measurements of quasar absorption line systems.  Composites
can also be made from sub-samples of the input spectra chosen according
to quasar properties such as luminosity, redshift, and radio loudness.
The dependence of global spectral characteristics on various quasar
properties will be the subject of a future paper \citep{vandenberk01}.
Here we concentrate on the continuum and emission line properties of
the global composite.   We describe the SDSS quasar sample in
\S\,\ref{sample}, and the method used to generate the composite spectra
in \S\,\ref{generate}.  The continuum and emission line features are
measured and discussed in \S\S\,\ref{features} and \ref{discussion}.
Wavelengths throughout the paper are vacuum values, except when using
the common notation for line identification (truncated air values for
wavelengths greater than $3000${\AA}, and truncated vacuum values for 
wavelengths less than $3000${\AA}).  We use the following values for
cosmological parameters throughout the paper: $H_0=100$\,km\,s$^{-1}$,
$\Omega_m=1.0$, $\Omega_\Lambda=0$, ($q_0=0.5$).

\section{The SDSS Quasar Sample\label{sample}}
The spectra were obtained as part of the commissioning phase of the
Sloan Digital Sky Survey.  Details of the quasar candidate target
selection and spectroscopic data reduction will be given in future
papers \citep{richards01b,newberg01,frieman01}.  The process is
summarized here.  Quasar candidates are selected in the color space of
the SDSS $u'g'r'i'z'$ filter system \citep{fukugita96}, from objects
found in imaging scans with the SDSS 2.5m telescope and survey camera
\citep{gunn98}.  The effective central wavelengths of the filters
for a power-law spectrum with a frequency index of $\alpha=-0.5$, are
approximately $3560, 4680, 6175, 7495, {\rm and} \,8875${\AA} for
$u'g'r'i'z'$ respectively.  Quasar candidates are
well-separated from the stellar locus in color space, and the filter
system allows the discovery of quasars over the full range of redshifts
from $z = 0$ to $z\approx 7$.  The locations of known quasars in the
SDSS color space as a function of redshift are shown by \citet{fan99,
fan00, fan01, newberg99, schneider01} and especially
\citet{richards01a} who plot the locations of over $2600$ quasars for
which there is SDSS photometry.  Quasar candidates are selected to
$i'\approx 19$ in the low-redshift ($z\lesssim 2.5$) regions of color
space, and no discrimination is made against extended objects in those
regions.  High-redshift quasar candidates are selected to $i'\approx
20$.  Objects are also selected as quasar candidates if they are point
sources with $i'\le19$ and match objects in the VLA FIRST radio source
catalog \citep{becker95}.  Thus, quasars in the SDSS are selected both
by optical and radio criteria.  These data were taken while the
hardware, and in particular the target selection software, was being
commissioned. Therefore, the selection criteria for quasars has evolved
somewhat over the course of these observations, and will not exactly
match the final algorithm discussed in \citet{richards01b}.
Because of the changing quasar selection criteria and the loose
definition of 'quasar', discretion should be exercised when using the
global composite spectra generated from this quasar sample as templates
for quasars in other surveys, or subsets of the SDSS quasar sample.

The candidates were observed using the 2.5m SDSS telescope
\citep{siegmund01} at Apache Point Observatory, and a pair of double
fiber-fed spectrographs \citep{uomoto01}. Targeted objects are grouped
into 3 degree diameter ``plates'', each of which holds 640 optical
fibers.  The fibers subtend 3$\arcsec$ on the sky and their positions
on the plates correspond to the coordinates of candidate objects, sky
positions, and calibration stars.  Approximately 100 fibers per plate
are allocated to quasar candidates.  At least three 15 minute exposures
are taken per plate.  So far, spectra have been taken mainly along a
$2.5$ degree wide strip centered on the Celestial Equator, with a
smaller fraction at other declinations.  The spectra in this study were
grouped on 66 plates which overlap somewhat to cover approximately
$320$ square degrees of sky covered by the imaging survey.  The plates
were observed from October, 1999 to June, 2000.  The raw spectra were
reduced with the SDSS spectroscopic pipeline \citep{frieman01}, which
produces wavelength- and flux-calibrated spectra that cover an
observed wavelength range from $3800 - 9200${\AA} at a spectral
resolution of approximately $1850$ blueward of $6000${\AA}, and 
2200 redward of $6000${\AA}.  These spectra and more will be made
publicly available -- in electronic form -- in June 2001 as part
of the SDSS Early Data Release \citep{stoughton01}.

The flux calibration is only approximate at this time, and a point
which deserves elaboration since it is the most important source of
uncertainty in the continuum shapes of the spectra.  Light losses from
differential refraction during the observations are minimized by
tracking guide stars through a $g'$ filter -- the bluest filter within
the spectral range.  Several F subdwarf stars are selected for
observation (simultaneously with the targeted objects) on each plate.
One of these -- usually the bluest one -- is selected, typed, and used
to define the response function.  This process also largely corrects
for Galactic extinction, since the distances to the F subdwarfs
employed are typically greater than $2.5$kpc, and all of the survey area
is at high Galactic latitude.  Uncertainties can arise in the spectral
typing of the star, and from any variation in response across a plate.
A check on the accuracy of the flux calibration is made for each plate
by convolving the calibrated spectra with the filter transmission
functions of the $g'$, $r'$, and $i'$ bands, and comparing the result
to magnitudes derived from the imaging data using an aperture the same
diameter as the spectroscopic fibers.  For a sample of about 2300 SDSS
quasar spectra, the median color difference between the photometric and
spectral measurements, after correcting the photometric values for
Galactic extinction \citep{schlegel98}, was found to be
$\Delta(g'-r')\approx 0.01$ and $\Delta(r'-i')\approx 0.04$.  This
means that the spectra tend to be slightly bluer than expectations from
photometry.  For a pure power-law spectrum with true frequency index of
$\alpha_{\nu}=-0.5$, which is often used to approximate quasars, the
difference in both colors would result in a measured index which is
systematically greater (bluer) by about $0.1$.  Quasar spectra are not
pure power-laws, and the color differences are well within the
intrinsic scatter of quasars at all redshifts \citep{richards01a}.
Also, the SDSS photometric calibration is not yet finalized, and the
shapes of the filter transmission curves are still somewhat uncertain,
both of which could contribute to the spectroscopic vs.\ photometric
color differences.  The colors of the combined spectra agree well with
the color-redshift relationships found by \citet{richards01a}, (see
\S\S\,\ref{emission},\ref{discussion}) which also leads us to believe
that the flux calibrations are reasonably good.  However, we caution
that the results here on the combined continuum shape cannot be
considered final until the SDSS spectroscopic calibration is verified.

\begin{figure}
\includegraphics[scale=0.7]{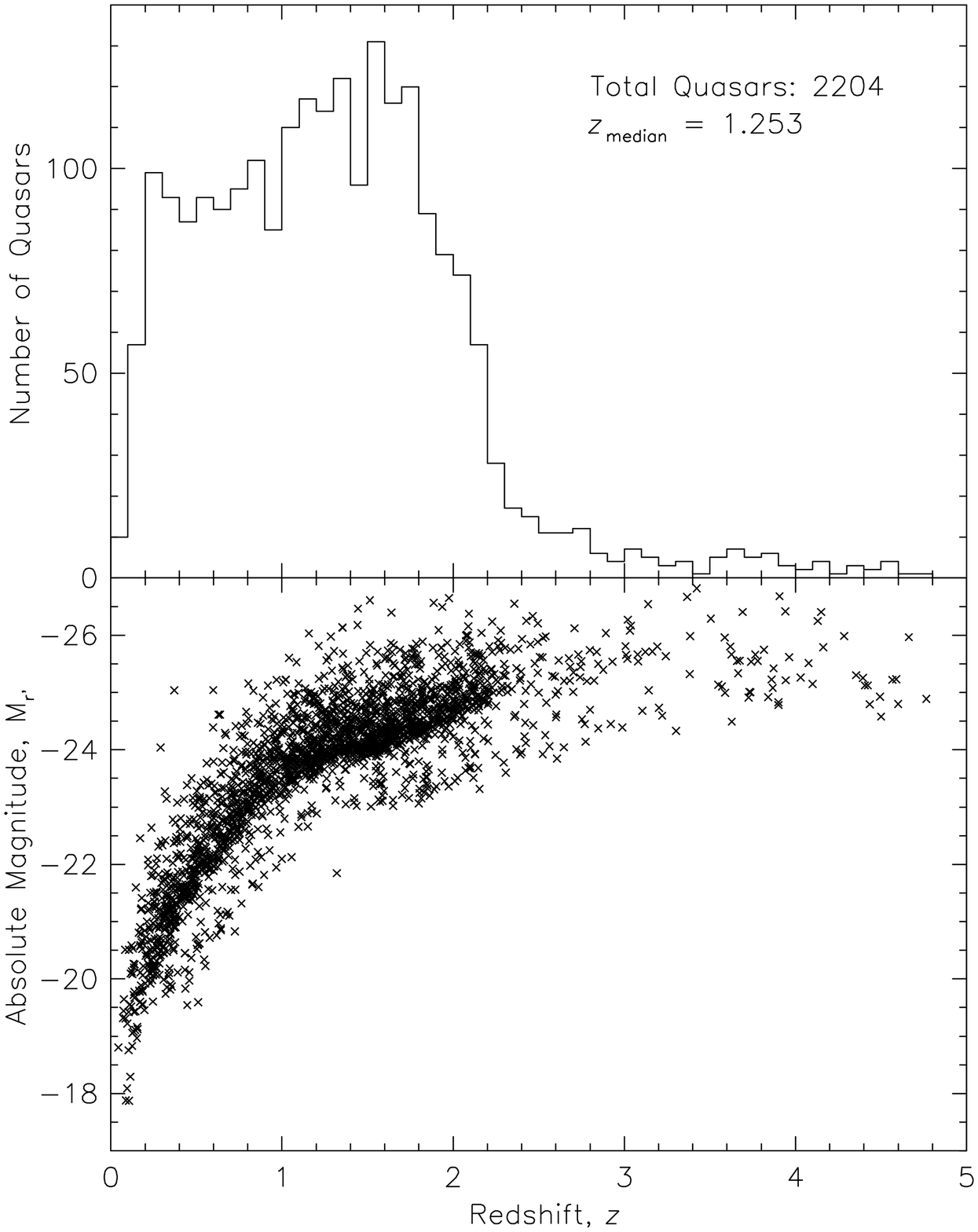}
\caption{Redshift distribution of the 2204 quasars used for the
composite spectra (top), and the absolute $r'$ magnitude, $M_{r'}$,
vs.\ redshift (bottom).
The median redshift is $z=1.253$. \label{zdist}}
\end{figure}

Quasars were identified from their spectra and approximate redshift
measurements were made by manual inspection.\footnote{Refined redshift
measurements were made later as described in \S\,\ref{generate}.} We
define quasar to mean any extra-galactic object with at least one broad
emission line, and that is dominated by a non-stellar continuum.  This
includes Seyfert galaxies as well as quasars, and we do not make a
distinction between them.  Spectra were selected if the rest-frame FWHM
of the strong permitted lines, such as \civ, \mgii, and the Balmer
lines, were greater than about $500$\,km\,s$^{-1}$.  In most cases
those line widths well exceeded $1000$\,km\,s$^{-1}$.  Since we require
only one broad emission line, some objects which may otherwise be
classified as ``Type 2'' AGN (those with predominantly narrow emission
lines) are also included in the quasar sample.  Spectra with continua
dominated by stellar features, such as unambiguous Ca~H and K lines, or
the $4308${\AA} G-band, were rejected.  This definition is free from
traditional luminosity or morphology-based criteria, and is also
intended to avoid introducing a significant spectral component from the
host galaxies (see \S\,\ref{discussion}).  Spectra with broad
absorption line features (BAL quasars), which comprise about $4\%$ of
the initial sample, were removed from the input list.  We are studying
BAL quasars in the SDSS sample intensively, and initial results are
forthcoming \citep[e.g.]{menou01}; the focus of the present paper is on
the intrinsic continua of quasars, and BAL features can heavily obscure
the continua.  Other spectra with spurious artifacts introduced either
during the observations or by the data reduction process (about $10\%$
of the initial sample) were removed from the input list.\footnote{These
artifacts are due to the inevitable problems of commissioning both the
software and hardware, and the problem rate is now negligible.} Spectra
obtained as part of SDSS follow-up observations on other telescopes,
such as the high-redshift samples of \citet{fan99, fan00, fan01},
\citet{schneider00, schneider01}, and \citet{zheng00} were not
included.  Figure\,\ref{zdist} shows the redshift distribution of the
quasars used in the composite, and the absolute $r'$ magnitudes
vs.\ redshift.   Discontinuities in the selection function for the
quasars, such as the fainter magnitude limit for high-redshift
candidates, are evident in Figure\,\ref{zdist}.  The final list of
spectra contains $2204$ quasars spanning a redshift range of $0.044
\leq z \leq 4.789$, with a median quasar redshift of
$\overline{z}=1.253$.  The vast majority of the magnitudes lie in the
range $17.5 < r' < 20.5$.

\section{Generating the Composites\label{generate}}
The steps required to generate a composite quasar spectrum involve selecting
the input spectra, determining accurate redshifts, rebinning the spectra
to the rest frame, scaling or normalizing the spectra, and stacking the
spectra into the final composite.  Each of these steps can have many
variations, and their effect on the resulting spectrum can be
significant (see \citet{francis91} for a discussion of some of these
effects).  The selection of the input spectra was described in
the previous section, and here we detail the remaining steps.

The appropriate statistical methods used to combine the spectra depend
upon the spectral quantities of interest. We are interested in both the
large-scale continuum shape and the emission line features of the
combined quasars.  We have used combining techniques to generate two
composite spectra: 1) the median spectrum which preserves the relative
fluxes of the emission features; and 2) the geometric mean spectrum
which preserves the global continuum shape.  We have used the geometric
mean because quasar continua are often approximated by power-laws, and
the median (or arithmetic mean) of a sample of power-law spectra will
not in general result in a power-law with the mean index.  The
geometric mean is defined as, $\langle f_{\lambda} \rangle_{gm} =
(\prod^{n}_{i=1}f_{\lambda,i})^{1/n}$, where $f_{\lambda,i}$ is the
flux density of spectrum number $i$ in the bin centered on wavelength
$\lambda$, and $n$ is the number of spectra contributing to the bin.
Assuming a power-law form for the continuum flux density, $f_{\lambda}
\propto \lambda^{-(\alpha_{\nu} + 2)}$, it is easily shown that
$\langle f_{\lambda} \rangle_{gm} \propto \lambda^{-(\langle
\alpha_{\nu} \rangle + 2)}$, where $\langle \alpha_{\nu} \rangle$ is
the (arithmetic) mean value of the frequency power-law index.  (The
wavelength index, $\alpha_\lambda$, and the frequency index,
$\alpha_\nu$, are related by $\alpha_\lambda = -(\alpha_{\nu} + 2)$.)

The rest positions of emission lines in quasar spectra, especially the
high-ionization broad lines,  are known to vary from their nominal
laboratory wavelengths \citep{gaskell82, wilkes86, espey89, zheng90,
corbin90, weymann91, tytler92, brotherton94b, laor95, mcintosh99}, so
the adopted redshifts of quasars depend upon the lines measured. In
addition to understanding the phenomenon of line shifts, unbiased
redshifts are important for understanding the nature of associated
absorption line systems \citep{foltz86}, for accurately measuring the
intergalactic medium ionizing flux \citep{bajtlik88}, and understanding
the dynamics of close pairs of quasars.  If the redshifts are
consistently measured, say using a common emission line or by
cross-correlation with a template, then the mean relative line shifts
can be measured accurately with a composite made using those
redshifts.  For the redshifts of our quasars, we have used only the
\oiii\,$\lambda 5007$ emission line when possible, since it is narrow,
bright, unblended, and is presumed to be emitted at nearly the systemic
redshift of the host galaxy \citep{gaskell82,vrtilek85,mcintosh99}.
Some weak Fe\,{\sc ii} emission is expected near $5000${\AA}
\citep[e.g.]{wills85,verner99,forster01}, but after subtraction of a
local continuum (see \S\,\ref{emission}), contamination of the
narrow \oiii line by the broad Fe\,{\sc ii} complex should be less
than a few percent at most.  Additionally, we use only the top
$\approx 50\%$ of the emission line peak to measure its position,
which greatly reduces uncertainties due to line asymmetry.
An initial composite was made (as described below) using spectra
with measured \oiii emission, and this composite was used as a
cross-correlation template for quasars in which the \oiii line was not
observable.  In this way, all quasars were put onto a common redshift
calibration, i.e. relative to the \oiii$\lambda 5007$ line.  We now
explain this in detail.

\subsection{Generating the [O\,{\sc iii}] Template\label{o3template}}

The \oiii based spectrum was made using $373$ spectra with a strong
\oiii\,$\lambda 5007$ emission line unaffected by night sky lines, and
includes quasars with redshifts from $z = 0.044$ to $z = 0.840$.
Spectra were combined at rest wavelengths which were covered by at
least 3 independent spectra, which resulted in a final wavelength
coverage of $2070{\rm {\AA}} < \lambda < 8555${\AA} for the \oiii based
spectrum.  The redshifts were based upon the peak position of the
\oiii\,$\lambda 5007$ line, estimated by calculating the mode of the
top $\approx 50\%$ of the line using the relation, ${\rm mode} =
3\times{\rm median} - 2\times{\rm mean}$, which gives better peak
estimates than the centroid or median for slightly skewed profiles
\citep[e.g.][]{lupton93}.  Uncertainties in the peak positions were
estimated by taking into account the errors in the flux density of the
pixels contributing to the emission line.  The mean uncertainty in the
peak positions was $35$\,km\,s$^{-1}$ (rest frame velocity).  This is a
few times larger than the wavelength calibration uncertainty of
$<10$\,km\,s$^{-1}$, based upon spectral observations of radial
velocity standards \citep{york00}.  The wavelength array of each
spectrum was shifted to the rest frame using the redshift based on the
\oiii line.  The wavelengths and flux densities were rebinned onto a
common dispersion of $1${\AA} per bin -- roughly the resolution of the
observed spectra shifted to the rest frame -- while conserving flux.
Flux values in pixels which overlapped more than one new bin were
distributed among the new bins according to the fraction of the
original pixel width covering each new bin.  The spectra were ordered
by redshift and the flux density of the first spectrum was arbitrarily
scaled.  The other spectra were scaled in order of redshift to the
average of the flux density in the common wavelength region of the mean
spectrum of all of the lower redshift spectra.  The final spectrum was
made by finding the median flux density in each bin of the shifted,
rebinned, and scaled spectra.

The \oiii based median composite was then used as a template to refine
redshift estimates for those spectra without measurable \oiii emission
and those for which the \oiii line was redshifted beyond
$9200$\,{\AA}.  We used a $\chi^{2}$ minimization technique, similar to
that used by \citet{franx89} to measure the redshifts.  A low-order
polynomial was fit to the composite and to each spectrum to approximate
a continuum, then subtracted.  The composite spectrum was shifted in
small redshift steps and compared to the individual quasar spectra.
The redshift which minimized $\chi^{2}$ -- the sum of the squared
inverse-variance weighted residuals -- was taken as the systemic
redshift.  Quasars with $\chi^2$ and manual redshifts which differed by
more than twice the dispersion of the velocity differences of the
entire sample (about $700$\,km\,s$^{-1}$) were examined for possible
causes unrelated to the properties of the quasar.  Spectra with
identified problems were either corrected (if possible) or rejected.  A
new composite was then made using all of the spectra with either
$\chi^2$ or \oiii redshifts.  The template matching and recombining
process was done in several progressively higher redshift ranges, so
that there was sufficient overlap between the templates and the input
spectra that included at least two strong emission lines.

\subsection{Generating the Composite Spectra\label{makecomposites}}

\begin{figure}
\includegraphics[scale=0.7,angle=-90]{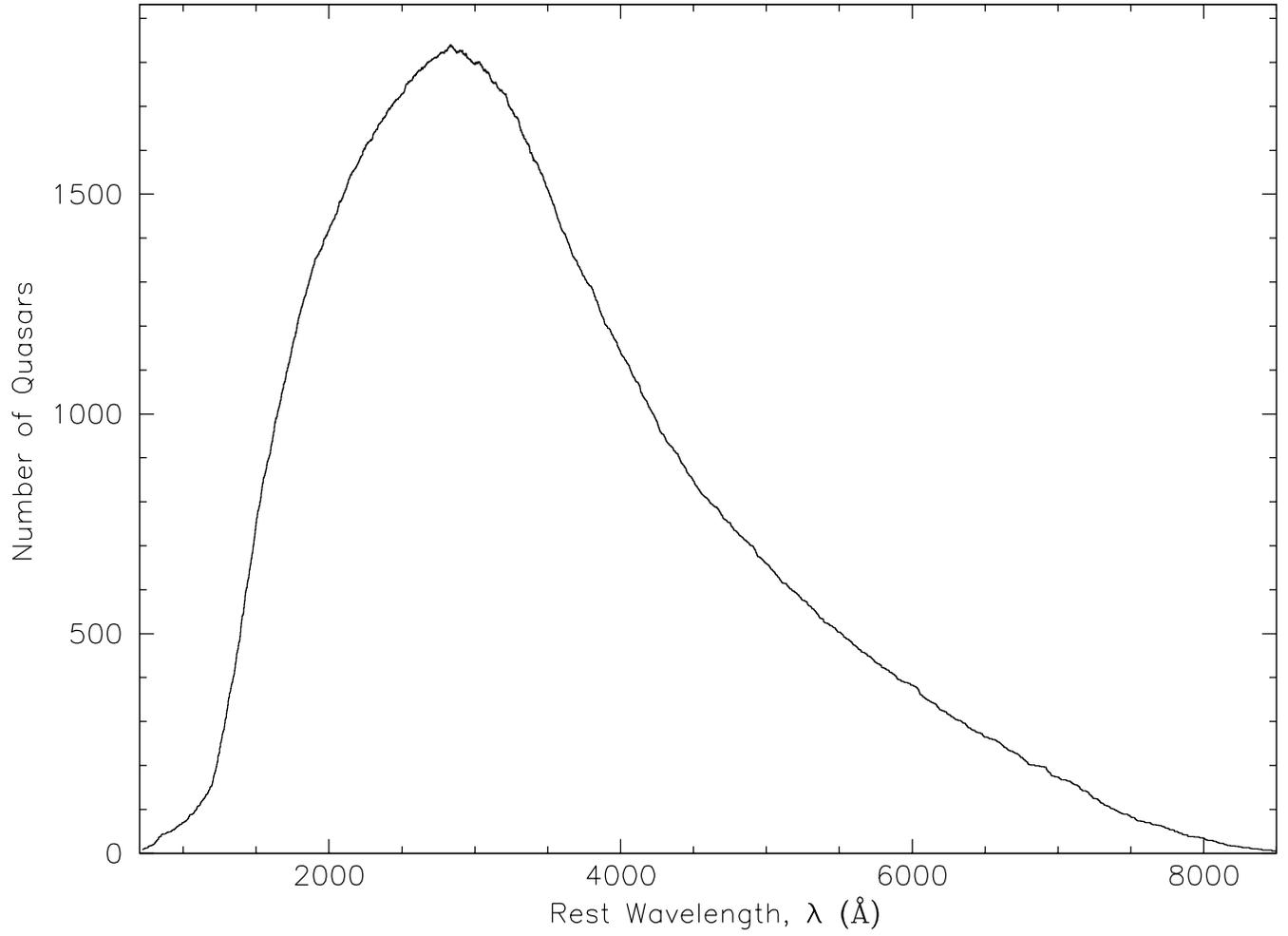}
\caption{Number of quasar spectra combined in each 1\,{\AA} bin of the composite
as a function of rest wavelength. \label{nqso}}
\end{figure}

\begin{figure}
\includegraphics[scale=0.7,angle=-90]{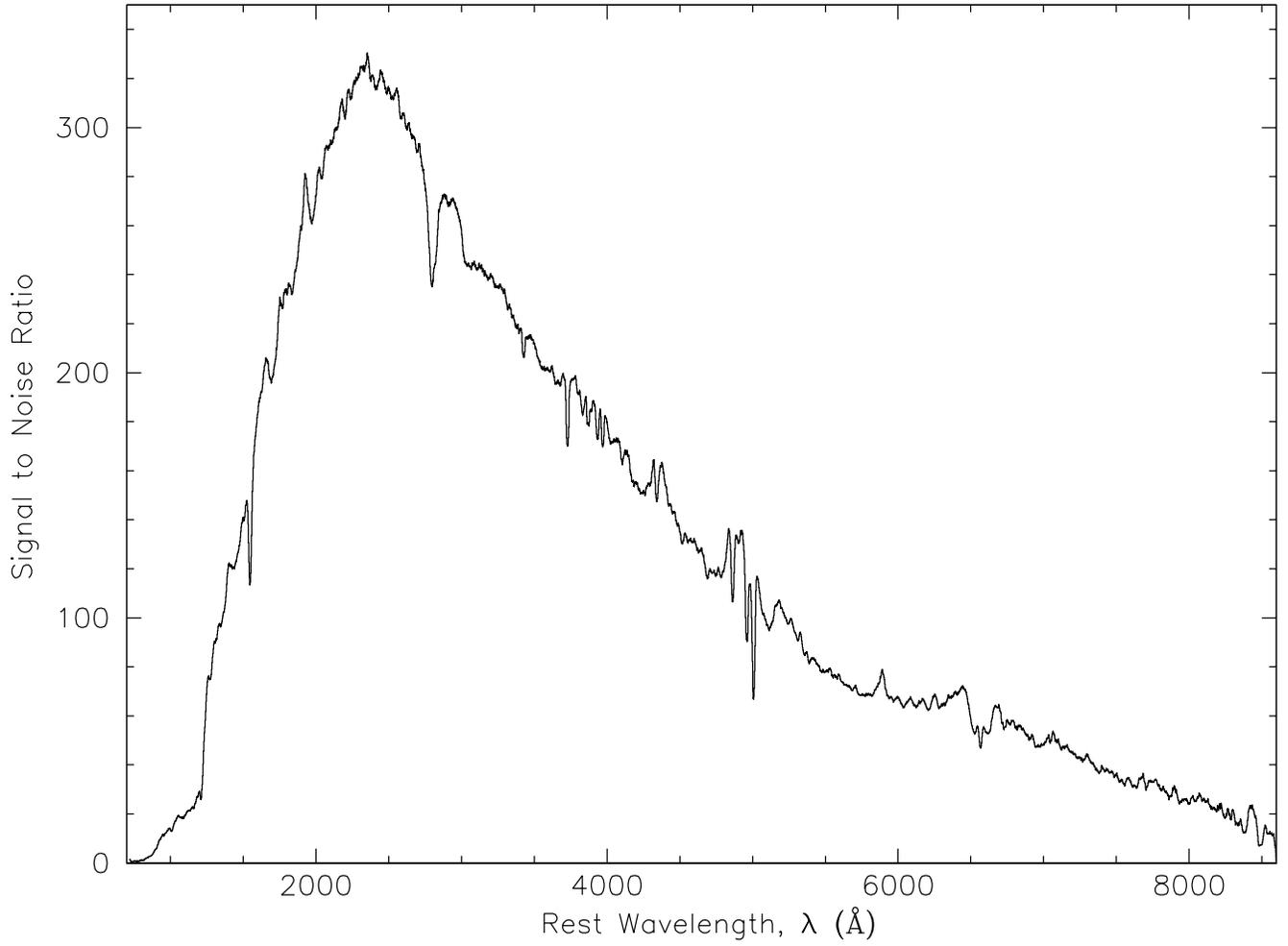}
\caption{Signal to noise ratio per $1${\AA} bin for the median composite
quasar spectrum.  The peak reaches almost 330 at $2800$\,{\AA}.
\label{sncomp}}
\end{figure}

Both median and geometric mean composite spectra were then generated
for the analysis of emission features and the global continuum
respectively.  The final set of spectra were shifted to the rest frame
using the refined redshifts, then rebinned onto a common wavelength
scale at 1\,{\AA} per bin, which is roughly the resolution of the
observed spectra shifted to the rest frame.  The number of quasar
spectra which contribute to each 1\,{\AA} bin is shown as a function of
wavelength in Fig.\,\ref{nqso}. The median spectrum was constructed
from the entire data set in the same way as the \oiii composite, as
described in the previous section.  The spectral region blueward of the
\lya emission line was ignored when calculating the flux density
scaling, since the \lya forest flux density varies greatly from
spectrum to spectrum.  The final spectrum was truncated to $800${\AA}
on the short wavelength end, since there was little or no usable flux
in the contributing spectra at shorter wavelengths.  The median flux
density values of the shifted, rebinned, and scaled spectra were
determined for each wavelength bin to form the final median composite
quasar spectrum, shown in Fig.\,\ref{lglgmed} on a logarithmic scale.
An error array was calculated by dividing the $68\%$ semi-interquantile
range of the flux densities by the square root of the number of spectra
contributing to each bin.  This estimate agrees well with the
uncertainty determined by measuring the variance in relatively
featureless sections of the combined spectrum.  The median spectrum
extends from $800 - 8555$\,{\AA} in the rest frame.
Figure\,\ref{sncomp} shows the S/N per 1\,{\AA} bin, which approaches
330 at 2800\,{\AA}.  The wavelength, flux density, and uncertainty
arrays of the median spectrum are given in Table\,\ref{medianspec},
which is available as an electronic journal table.

To generate the geometric mean spectrum, the shifted and rebinned
spectra were normalized to unit average flux density over the rest wavelength
interval $3020 - 3100$\,{\AA}, which contains no strong narrow emission lines,
and which is covered by about $90\%$ of the spectra.
The restriction that the input spectra cover this interval results in a
combined spectrum which ranges from about $1300 - 7300${\AA}, and
is composed of spectra with redshifts from $z= 0.26 - 1.92$.
The geometric mean of the flux density values was calculated in each
wavelength bin to form the geometric mean composite quasar spectrum,
shown in Fig.\,\ref{lglggm} on a logarithmic scale.
The median and geometric mean composites are quite similar, but there
are subtle differences in both the continuum slopes and the emission
line profiles, discussed further in the next sections, which justify
the construction of both composite spectra.

\section{Continuum, Emission, and Absorption Features \label{features}}

\subsection{The Continuum \label{continuum}}
\begin{figure}
\includegraphics[scale=0.7,angle=-90]{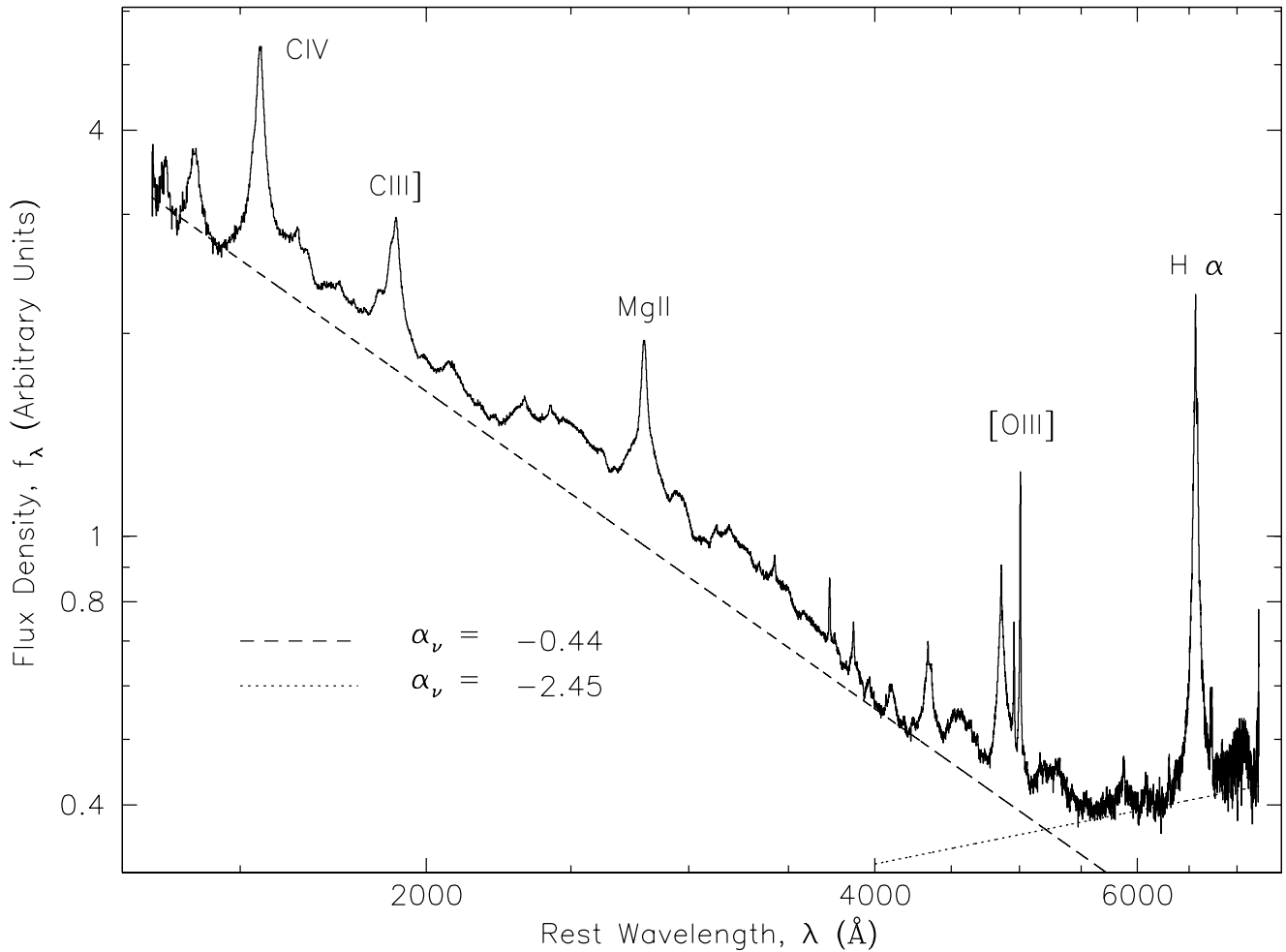}
\caption{Composite quasar spectrum generated using the geometric mean of the
input spectra.  Power-law fits to the estimated continuum flux are shown.
The geometric mean is a better estimator than the arithmetic mean
(or median) for power-law distributions. The resolution of the
input spectra is $\approx 1800$ in the observed frame, which gives
a wavelength resolution of about $1${\AA} in the rest frame. \label{lglggm}}
\end{figure}

The geometric mean spectrum is shown on a log-log scale in
Fig.\,\ref{lglggm}, where a single power-law will appear as a straight
line.  The problem of fitting the quasar continuum is complicated by
the fact that there are essentially no emission-line-free regions in
the spectrum.  Our approach is to find a set of regions which give the
longest wavelength range over which a power law fit does not cross the
spectrum (i.e. the end points of the fit are defined by the two most
widely separated consecutive intersections).  The regions which satisfy
this are $1350 - 1365$\,{\AA} and $4200 - 4230$\,{\AA}.  A single
power-law fit through the points in those regions gives an index of
$\alpha_{\nu} = -0.44 \phn (\alpha_{\lambda} = -1.56)$, and fits the
spectrum reasonably well from just redward of \lya to just blueward of
H\,$\beta$ (Fig.\,\ref{lglggm}).  The statistical uncertainty in the
spectral index from the fit alone is quite small ($\approx 0.005$)
owing to the high S/N of the spectrum and the wide separation of the
fitted regions.  However, the value of the index is sensitive to the
precise wavelength regions used for fitting.  More importantly, the
spectrophotometric calibration of the spectra introduces an uncertainty
of $\approx 0.1$ in $\alpha_{\nu}$ (\S\,\ref{sample}).  We estimate the
uncertainty of the measured value of the average continuum index to be
$\approx 0.1$, based mainly on the remaining spectral response
uncertainties.  Redward of H\,$\beta$ the continuum flux density rises
above the amount predicted by the power-law; this region is better fit
by a separate power-law with an index of $\alpha_\nu = -2.45 \phn
(\alpha_{\lambda} = 0.45)$ (Fig.\,\ref{lglggm}), which was determined
using the wavelength ranges $6005 - 6035${\AA} and $7160 - 7180${\AA}.
The abrupt change in the continuum slope is discussed in
\S\,\ref{discussion}.

\begin{figure}
\includegraphics[scale=0.7,angle=-90]{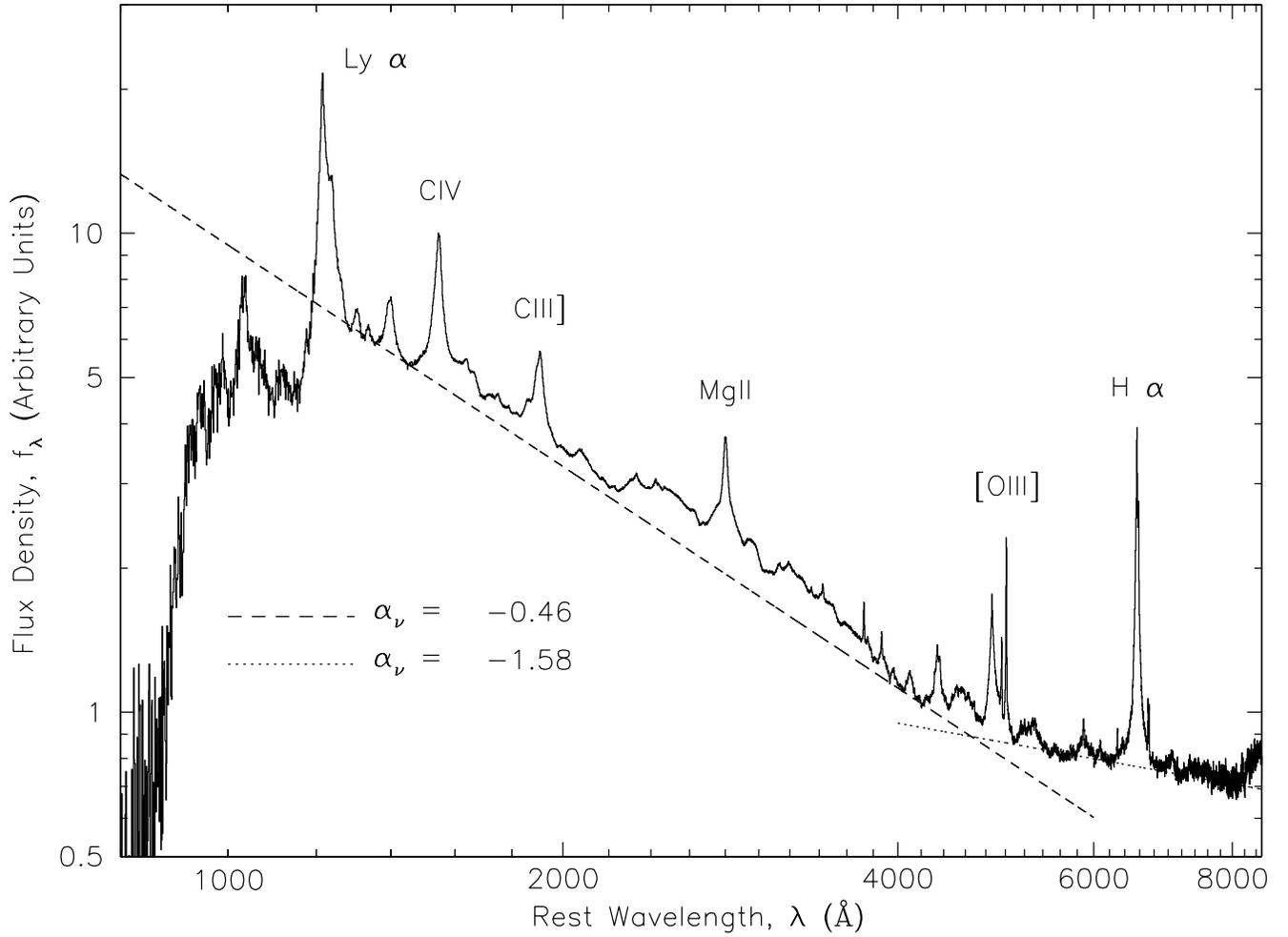}
\caption{Composite quasar spectrum using median combining.  Power-law fits
to the estimated continuum flux are shown.  The resolution of the
input spectra is $\approx 1800$, which gives a wavelength resolution
of about $1${\AA} in the rest frame. \label{lglgmed}}
\end{figure}

As a comparison, we have also measured the power-law indices for the
{\em median} composite, which are $\alpha_\nu = -0.46 \phn
(\alpha_{\lambda} = -1.54)$ and $\alpha_\nu = -1.58 \phn
(\alpha_{\lambda} = -0.42)$ for the respective wavelength regions
(Fig.\,\ref{lglgmed}).  The index found for the \lya to H\,$\beta$
region is almost indistinguishable from that found for the geometric
mean composite.  The indices for both spectra redward of H\,$\beta$ are
significantly different however, and are a result of the different
combining processes.  The geometric mean should give a better estimate
of the average index, but comparison with mean or median composite
spectra from other studies is probably reasonable in the \lya to
H\,$\beta$ region, given the small difference in the indices measured
for our composite spectra.  The continuum blueward of the \lya emission
line is heavily absorbed by \lya forest absorption, as seen in
Fig.\,\ref{lglgmed}.  However, because the strength of the \lya forest
is a strong function of redshift, and a large range of redshifts was
used in constructing the sample, no conclusions can be drawn about the
absorption or the continuum in that region.

\subsection{Emission and Absorption Lines \label{emission}}
\begin{figure}
\epsscale{0.8}
\plotone{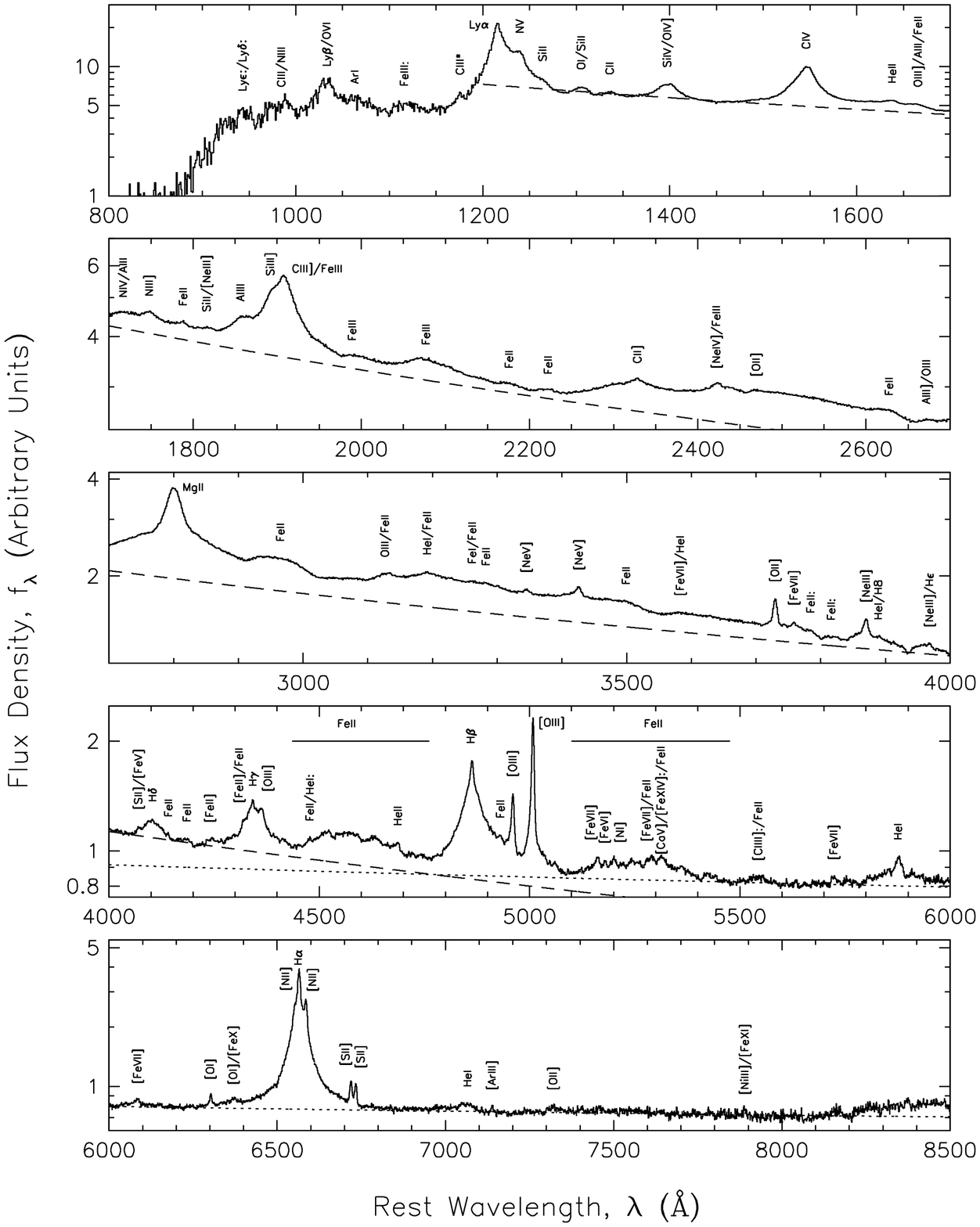}
\caption{Expanded view of median quasar composite on log-linear scale, with 
emission features labeled by ion.  Labels ending with a colon (:) are
uncertain identifications.  The two power-law continuum fits are shown
by dashed and dotted lines. The flux from 1600-3800\,{\AA} is also composed
of heavily blended Fe\,{\sc ii} and Fe\,{\sc iii} lines, and Balmer continuum 
emission. \label{complab}}
\end{figure}

The high S/N and relatively high resolution (1\,{\AA}) of the composite
allows us to locate and identify weak emission features and resolve
some lines that are often blended in other composites.  It is also
possible that our sample includes a higher fraction of spectra with
narrower line profiles, which could also help in distinguishing
emission features.  For example,
close lines that are clearly distinguished in the spectrum include
H$\alpha$/[N\,{\sc ii}] ($\lambda\lambda 6548, 6563, 6583$), Si\,{\sc
iii}]/C\,{\sc iii}] ($\lambda\lambda 1892, 1908$), the [S\,{\sc ii}]
($\lambda\lambda 6716, 6730$) doublet, and H$\gamma$/[O\,{\sc iii}]
($\lambda\lambda 4340, 4363$).  Emission line features above the
continuum were identified manually in the median spectrum.  Including
the broad \feii and \feiii complexes, $85$ emission features were
detected.  The end points of line positions, $\lambda_{lo}$ and
$\lambda_{hi}$, were estimated to be where the flux density was
indistinguishable from the local ``continuum''.  The local continuum is
not necessarily the same as the power-law continuum estimated in
\S\,\ref{continuum}, since the emission lines may appear to lie on top
of other emission lines or broad Fe\,{\sc ii} emission features.  The
peak position of each emission line, $\lambda_{obs}$, was estimated by
calculating the mode of the top $\approx 50\%$ of the line -- the same
method used for measuring the \oiii\,$\lambda 5007$ line peaks in
\S\,\ref{o3template}.  Uncertainties in the peak positions include the
contribution from the flux-densisty uncertainties, but none from
uncertainties in the local continuum estimate.  Fluxes and equivalent
widths were measured by integrating the line flux density between the
end points and above the estimated local continuum.  Line profile
widths were estimated by measuring the rms wavelength dispersion,
$\sigma_\lambda$, about the peak position -- i.e.\ the square-root of
the average flux-weighted squared differences between the wavelength of
each pixel in a line profile and the peak line position.  Asymmetry of
the line profiles was measured using Pearson's skewness coefficient,
${\rm skewness} = 3\times({\rm mean - median})/\sigma_\lambda$.  Lines
were identified by matching wavelength positions and relative strengths
of emission features found in other objects, namely the
\citet{francis91} composite, the \citet{zheng97} composite, the
narrow-lined quasar I~Zw~1 \citep{laor97, oke79, phillips76}, the
ultra-strong \feii emitting quasar 2226-3905 \citep{graham96}, the
bright Seyfert 1 galaxy NGC~7469 \citep{kriss00}, the high-ionization
Seyfert 1 galaxy III~Zw~77 \citep{osterbrock81}, the extensively
observed Seyfert 2 galaxy NGC~1068 \citep{snijders86}, the powerful
radio galaxy Cygnus~A \citep{tadhunter94}, and the Orion Nebula HII
region \citep{osterbrock92}.  Identification of many \feii complexes
was made by comparison with predicted multiplet strengths by
\citet{verner99, netzer83, grandi81}; and \citet{phillips78}, and
multiplet designations are taken from those references.
Table\,\ref{emlines} lists the detected lines, their vacuum wavelength
peak positions, relative fluxes, equivalent widths, profile widths,
skewness, and identifications.  Rest wavelengths were taken from the
Atomic Line List\footnote{The Atomic Line List is hosted by the
Department of Physics and Astronomy at the University of Kentucky; {\tt
(http://www.pa.uky.edu/$\sim$peter/atomic/)}.}.  Wavelengths of lines
consisting of multiple transitions were found by taking the
oscillator-strength weighted average in the case of permitted lines,
and the adopted values from the above references for forbidden lines.
In all cases, the permitted rest wavelength values agreed with the
(vacuum) values adopted in the above references.  Fig.\,\ref{complab}
shows an expanded view of the quasar composite on a log-linear scale
with the emission features labeled.

\begin{figure}
\epsscale{0.8}
\plotone{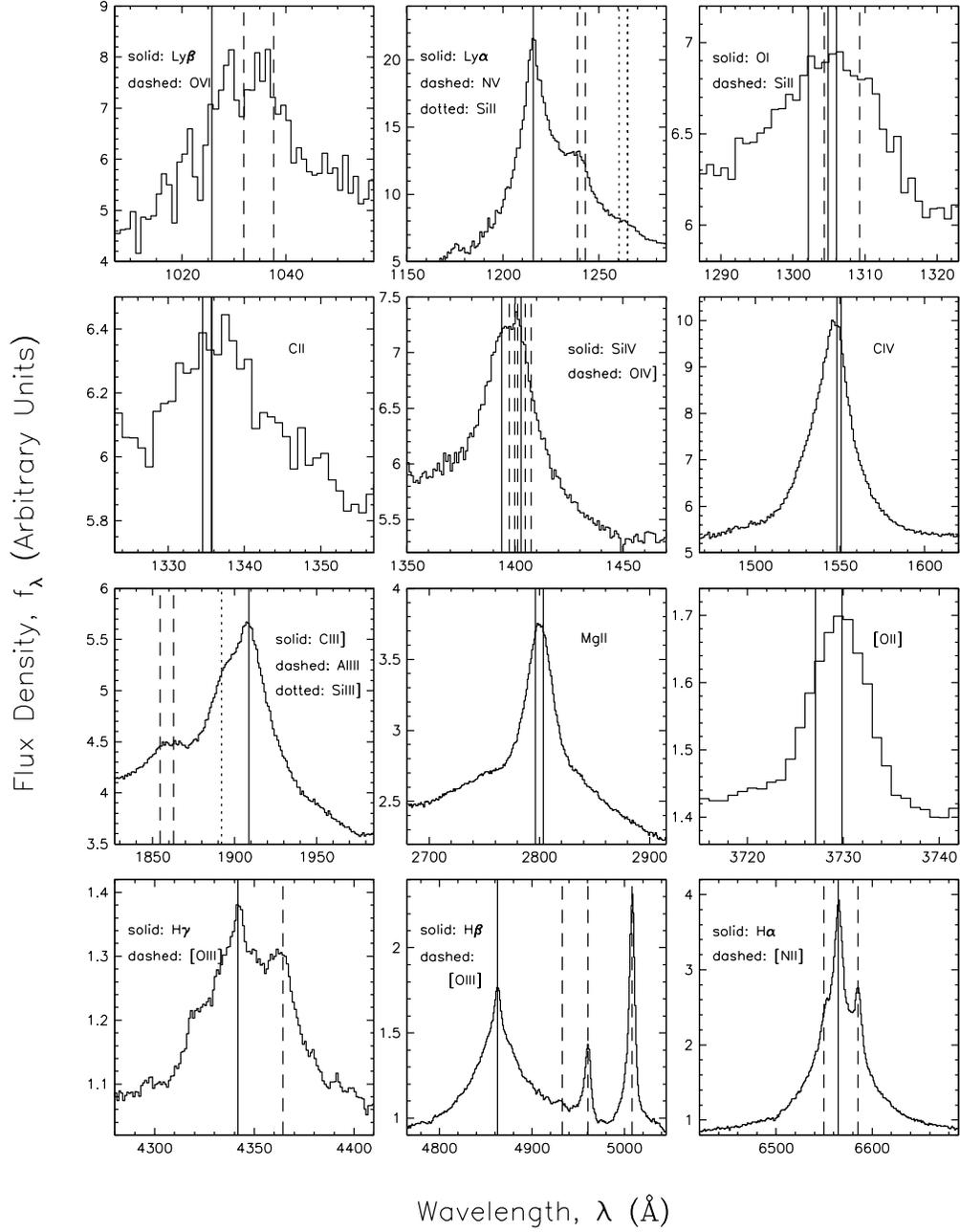}
\caption{Detailed view of 12 strong emission line regions.  The laboratory
rest wavelength positions of the major line components are shown.  Many of
the emission lines are composed of blended multiple transitions (doublets,
triplets, etc.) from the same ion.
\label{emclose}}
\end{figure}

It is clear from Fig.\,\ref{complab} that most of the UV-optical
continuum is populated by emission lines.  Most strong emission lines
show ``contamination'' by blends with weaker lines, as seen in the
expanded profiles of 12 emission line regions in Fig.\,\ref{emclose}.
The very broad conspicuous feature from $\approx 2200 - 4000$\,{\AA} is
known as the $3000$\,{\AA} bump \citep{grandi82, shields84}, and
consists of blends of \feii line emission and Balmer continuum
emission \citep{wills85}.  The \feii and \feiii complexes are particularly
ubiquitous, and contribute a large fraction of the emission line flux.
Using this composite, these complexes have been shown to be an important
contributor to the color-redshift relationships of quasars
\citep{richards01a}.  

\begin{figure}
\epsscale{0.8}
\plotone{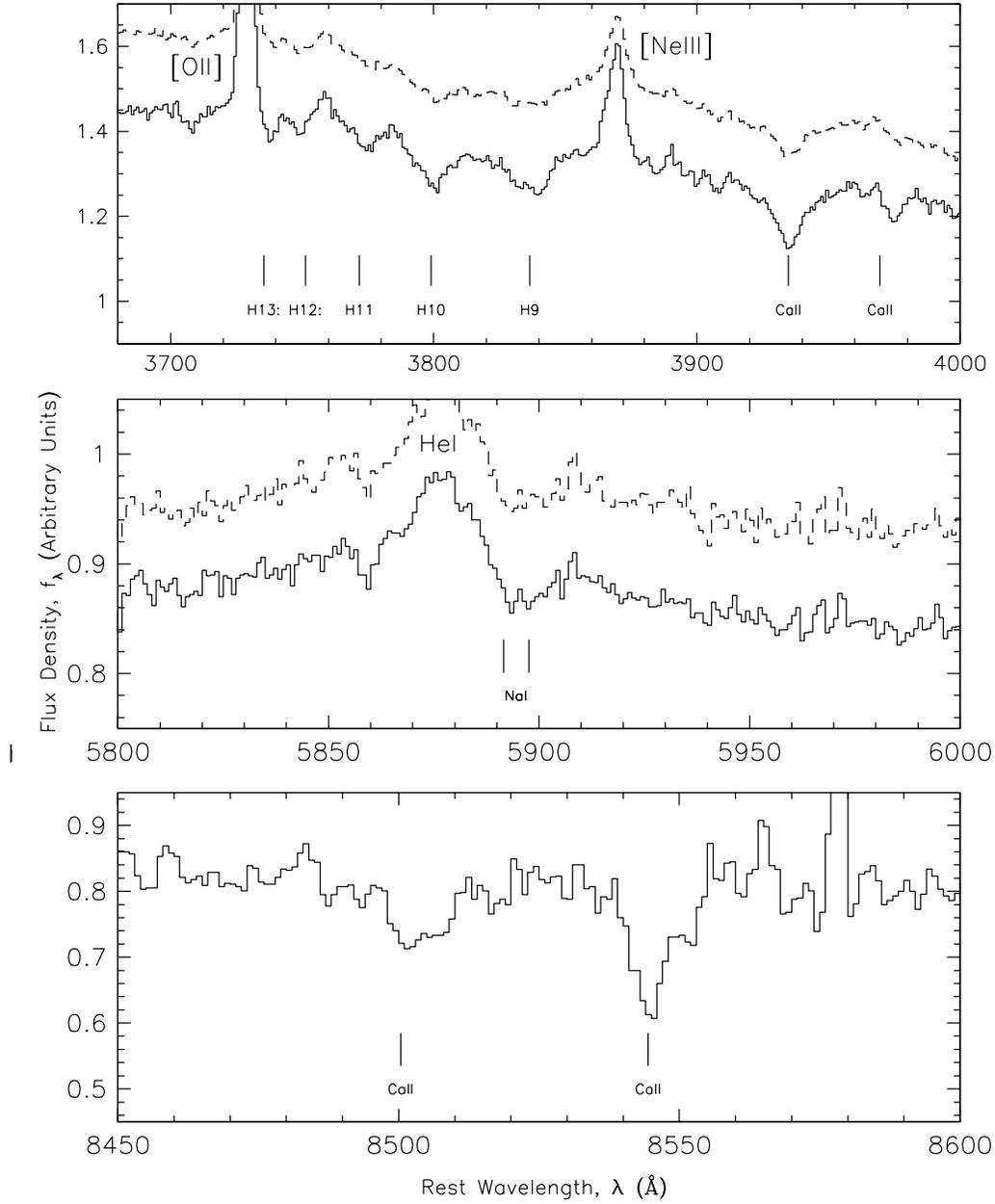}
\caption{Detailed view of absorption line regions in the low-redshift
composite quasar spectrum (solid line).  The laboratory rest wavelength
positions of detected absorption lines are labeled by ion.  Several
strong emission lines are also labeled.  The Ca\,{\sc ii}$\lambda 3968$
line is contaminated by emission from [Ne\,{\sc iii}]$\lambda 3967$ and
H\,$\epsilon$.  The full-sample composite spectrum (dashed line, offset)
is also shown for comparison in the top two panels (the spectra
are identical in the wavelength region covered in the bottom panel).
\label{lowzmed}}
\end{figure}

Several absorption features often seen in galaxies are also
identifiable in the median composite quasar spectrum.  These lines are
listed in Table\,\ref{ablines} along with several measured quantities,
and include H\,9 $\lambda 3835$, H\,10 $\lambda 3797$, the Ca\,{\sc ii}
$\lambda 3933$ K line, and Ca\,{\sc ii} $\lambda\lambda 8498, 8542$ --
two lines of a triplet (the second-weakest third component would fall
beyond the red end of the spectrum).  The Mg{\sc I} $\lambda\lambda
5167, 5172, 5183$ triplet lines may also be present in the spectrum,
but they would lie inside a strong complex of Fe{\sc ii} emission and
near several other expected emission lines.  The locations of other
common stellar absorption lines seen in galaxies, such as the
lower-order Balmer lines and the Ca\,{\sc ii} $\lambda 3968$ H line,
are dominated by emission lines.  The presence of stellar absorption
lines argues for at least some host galaxy contamination in the quasar
composite spectrum, despite the fact that we rejected objects with
obvious stellar lines in individual galaxies.  To examine this further,
we have created a low-redshift median composite using only quasars with
redshifts $z_{em} \leq 0.5$, which is almost equivalent to selecting
only quasars with restframe absolute $r'$ magnitude $M_{r'} \geq -21.5$
(calculated using a spectral index of $\alpha_\nu=-0.44$).  The low-$z$
composite covers a rest wavelength range of $2550 - 8555${\AA}.  The
absorption lines found in the low-$z$ spectrum are marked in
Fig.\,\ref{lowzmed}, and listed in Table\,\ref{ablines}.  More
absorption lines are detected in the low-$z$ composite spectrum than
the full-dataset spectrum, and the lines in common are stronger in the
low-$z$ spectrum -- as expected if host galaxy contamination is the
source of the absorption lines.  We discuss the absorption lines in
more detail in \S\,\ref{discussion}.

The $2175${\AA} extinction bump often seen in the spectra of objects
observed through the Galactic diffuse interstellar medium, and usually
attributed to graphite grains \citep{mathis77}, is not present at a
detectable level in the composite spectrum.  This agrees with the
non-detection of the feature by \citet{pitman00} who searched for it in
other quasar spectral composites.

\subsection{Systematic Line Shifts \label{shifts}}
Because the composite was constructed using redshifts based upon a
single emission line position (\oiii $\lambda 5007$) or
cross-correlations with an [O\,{\sc iii}]-based composite, we can check
for systematic offsets between the measured peak positions and the
[O\,{\sc iii}]-based wavelengths.  Several emission lines -- \civ
$\lambda 1549$ for example -- are offset from their laboratory
wavelengths, as evident from Fig.\,\ref{emclose}.  Such line shifts
have been detected previously
\citep[e.g.][]{grandi82,wilkes86,tytler92,laor95,mcintosh99}, and are
present for many of the lines listed in Table\,\ref{emlines}.  Real
line position offsets can be confused with apparent ``shifts'' which
can arise from several sources, including contamination by line blends,
incorrect identifications, and line asymmetry.  To minimize these
problems, we have selected only relatively strong lines with isolated
peaks, and we have re-measured the peaks of only the top 25\% of the
line flux for lines which appeared to have a very broad component or
asymmetric profile.  The velocity shifts for the selected lines are
listed in Table\,\ref{velshift}.  A negative velocity indicates that a
line is blueshifted with respect to the nominal laboratory wavelength,
and visa-versa.  By design, the \oiii$\lambda 5007$ line peak shows no
shift from its laboratory rest wavelength to well within the
measurement uncertainty.  All other emission line peaks are measured
with respect to the rest frame of the \oiii$\lambda 5007$ line.  The
two other measurable \oiii lines, $\lambda 4363$ and $\lambda 4958$,
have no velocity shift to within the uncertainties.

\begin{figure}
\epsscale{0.8}
\plotone{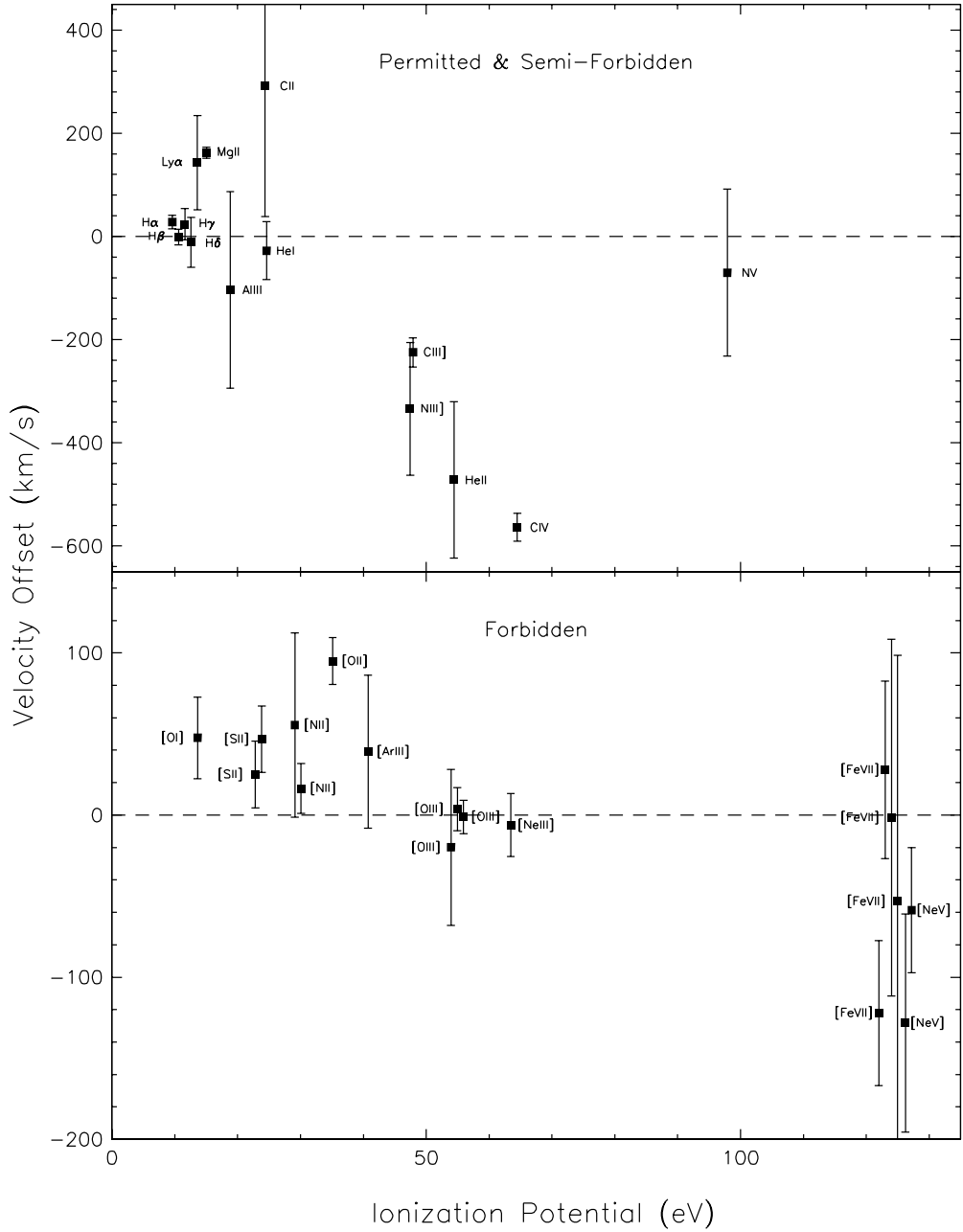}
\caption{Emission line velocity offsets relative to laboratory rest wavelengths
as a function of ionization potential for selected emission lines.  Error
bars show the 1 standard deviation uncertainty in the velocity measurement.
The points are labeled by ion.  Ionization potentials  corresponding to the
same ion are slightly offset from each other for clarity.  Permitted and
semi-forbidden lines are shown in the top panel, and forbidden lines are
shown in the bottom panel.  \label{velion}}
\end{figure}

It has been suggested that there is a correlation between the line
shift and the ionization energy of the species
\citep[e.g.][]{tytler92,mcintosh99}.  Quasar emission lines are
generally separated into two broad categories:  the permitted and
semi-forbidden lines which are typically broad ($FWHM >
500$\,km\,s$^{-1}$), and the much narrower forbidden lines.  These
classes of lines are thought to arise from physically distinct
regions:  the parsec-scale Broad Line Region (BLR) and the
kiloparsec-scale Narrow Line Region (NLR) respectively.  Since their
origins are likely to be different, we treat the BLR and NLR lines
separately.  Figure\,\ref{velion} shows the ionization energy
vs.\ velocity shifts in Table\,\ref{velshift} both for the BLR lines,
and the NLR lines, labeled by their ions.  In both cases, there is an
apparent anti-correlation between the velocity shifts and ionization
potential in Fig.\,\ref{velion}.  The Spearman rank correlation
coefficient for the BLR lines gives a random probability of finding as
strong a correlation at about $0.6\%$.  We have taken the uncertainties
in the velocity measurements into account, by creating $10^4$ mock data
sets of velocities, randomly distributed for each emission line
according to the velocity uncertainties, then recalculated the
correlation probabilities.  For half of the mock datasets, the random
probability of a correlation was less than $1.6\%$ for the BLR lines.
Thus we would have found a significant anti-correlation between the
velocity offsets and ionization potentials for a majority of
independent measurements.  The low-ionization \cii\,$\lambda 1335$ line
has the maximum redshift at $292$\,km\,s$^{-1}$, and the
high-ionization \civ\,$\lambda 1549$ line has the maximum blueshift at
$-564$\,km\,s$^{-1}$.  The \nv point appears to be somewhat of an
outlier, possibly due to severe blending with \lya.  It is also
interesting that \nv does not follow the Baldwin effect
\citep{espey99}, the strength of which is otherwise anti-correlated
with ionization potential.  In any case, the rank correlation of the
velocity offset and ionization potential is not significantly stronger
when \nv it is removed, and we have no compelling reason to do so.
 The velocity offsets are not as strong for the NLR lines -- $\lesssim
100$\,km\,s$^{-1}$ -- but the Spearman rank correlation probability is
$1.3\times 10^{-4}$, which is quite significant, and we find the
probability is less than $1\%$ for half of the mock data sets.  We
discuss emission line velocity shifts further in \S\,\ref{discussion}.

\subsection{Spectrum-to-spectrum Differences \label{variation}}
\begin{figure}
\includegraphics[scale=0.7,angle=-90]{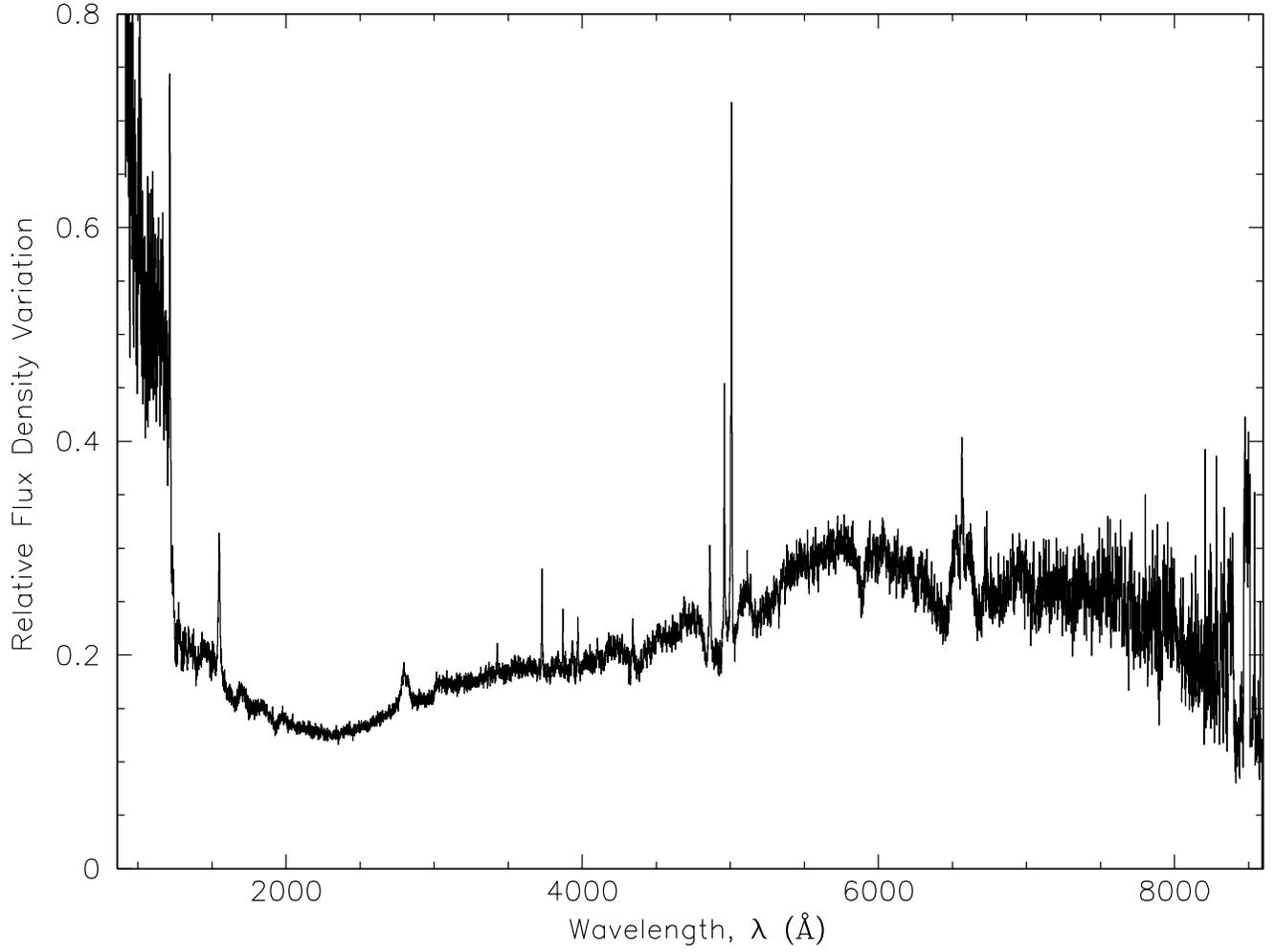}
\caption{Spectrum-to-spectrum variation of the quasar composite flux
density relative to the median flux as a function of rest
wavelength. \label{cov}}
\end{figure}

While constructing the median composite, the flux levels of overlapping
spectra were scaled so that the integrated flux densities were the
same.  Thus we expect the variation in the continuum flux density
across the spectrum to reflect the spectrum-to-spectrum differences
caused by differing continuum shapes and emission line fluxes and
profiles. (This does not, however, address spectral {\em time} variability.)
Fig.\,\ref{cov} shows the 68\% semi-interquantile range
divided by the median spectrum, after the contribution from the
combined flux density uncertainties of each spectrum were removed in
quadrature.  The individual spectral uncertainties include statistical
noise estimates, but not uncertainties in the (un-finalized) flux
calibration.  The largest relative variations from the median spectrum
occur in the narrow emission lines such as the \oiii $\lambda\lambda
4958, 5007$ lines, and the cores of broad emission lines such as
\civ$\lambda 1549$ and \lya$\lambda 1215$.  Variations of the broad
components of \halpha$\lambda 6563$ and \mgii$\lambda 2798$ are
evident, but less so for \hbeta$\lambda 4861$ and \civ$\lambda 1549$,
and there is little sign of variation in the \ciii$\lambda 1908$ line.
Most of the broad \feii complexes show significant variation.  The \lya
forest region varies considerably, as expected, since structure in the
forest can be partly resolved in the individual spectra, and the forest
strength changes with redshift so the combination of spectra at
different redshifts will naturally give rise to a high variance.  An
additional feature of some interest is the pair of variation peaks at
$3935$\,{\AA} and $3970$\,{\AA}, which correspond precisely to the
Ca\,{\sc ii} doublet.  These are detected in absorption in the median
composites (although [Ne\,{\sc iii}] and H\,$\epsilon$ emission
interfere with Ca\,{\sc ii}$\lambda 3970$), and the variation may
indicate that spectral contamination by the host galaxy is fairly
common.  A full analysis of spectrum-to-spectrum variations requires
other means such as principal component analysis \citep{boroson92,
francis92, brotherton94a,wills99}, which we plan for a future project.

\section{Discussion \label{discussion}}
\begin{figure}
\includegraphics[scale=0.7,angle=-90]{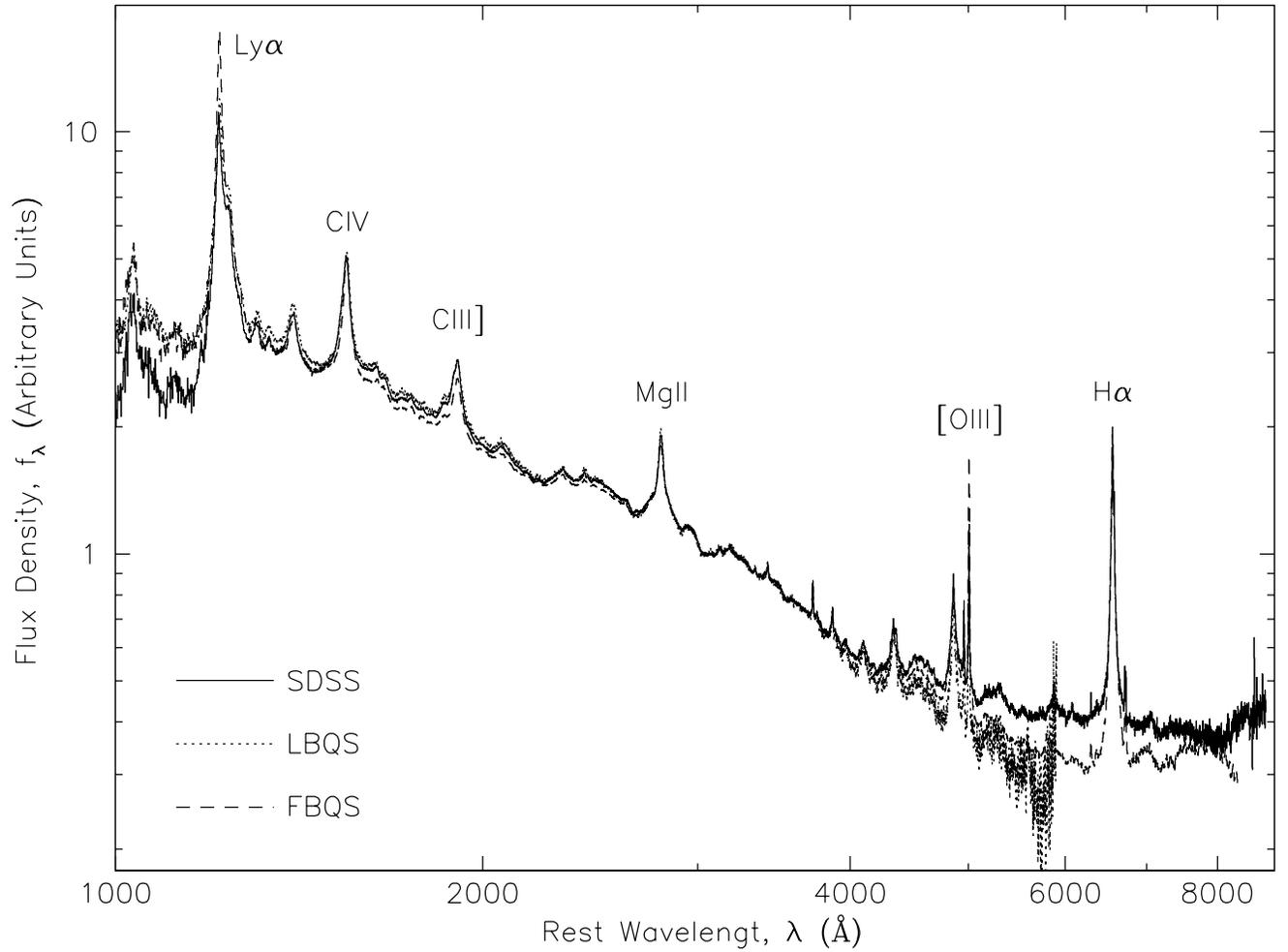}
\caption{Comparison of the SDSS median quasar composite spectrum (solid) with 
the LBQS (dotted) and FBQS (dashed) composites.  The spectra are scaled
to the same average flux density between $3020$ and $3100${\AA}.  Several
major emission lines are labeled for reference.  
\label{compare}}
\end{figure}

The Large Bright quasar Survey (LBQS) composite (kindly provided by
S.\ Morris) updated from \citet{francis91}, the First Bright quasar
Survey (FBQS) composite (available electronically,
\citet{brotherton00}), and our median composite, are shown for
comparison in Fig.\,\ref{compare}.  The spectra have been scaled to
unit average flux density in the range $3020 - 3100${\AA}.  All three
spectra are quite similar in appearance except for slight differences.
The strength of the \lya line and some of the narrow emission lines in
the FBQS composite are stronger than for the other composites.  The
difference is probably due to that fact that the FBQS sample is
entirely radio selected, and there is a correlation between line
strengths and radio loudness
\citep{boroson92,francis92,brotherton94a,wills99}.  Otherwise, the
relative fluxes are similar for the lines in common among the various
composites.  The higher resolution and higher S/N of our composite has
allowed us to identify many more lines than listed for the other
spectra (although a number of the features we find are present at a
lower significance level in the other spectra).  We have identified a
total of 85 emission features in the median spectrum.  All of the
features have been identified in other quasar or AGN spectra, but not
in any single object.  A large number of the identified features are
attributed to either \feii or \feiii multiplets.  The combination of
these features has been shown to greatly affect the color-redshift
relationship for quasars \citep{richards01a}.

A single power-law is an adequate fit to the continuum between \lya and
\hbeta, especially given the predicted strengths of the \feii and
\feiii emission line complexes in that range \citep{verner99,
netzer83,laor97}.  The index we find, $\alpha_\nu=-0.44$, is in good
agreement with most recent values found in optically-selected quasar
samples.  Table\,\ref{indices} lists average power-law indices from
various sources over the past decade. 
The LBQS composite, FBQS composite, and Hubble Space Telescope (HST)
composite \citep{zheng97}
spectra are available electronically,
so for consistency, we have also remeasured the power-law indices of
those spectra using the technique described in \S\,\ref{continuum}.
The remeasured values are not significantly different from the values
given in the papers.

Most of the composite measurements agree with averages over continuum
fits to individual spectra.  One outlier is the measurement by
\citet{zheng97}, who find a steeper (redder) continuum with
$\alpha_\nu=-0.99$ using a composite made with spectra from HST.  The
difference is attributed to the lower redshift of the \citet{zheng97}
quasar sample and a correlation between redshift and steeper UV
continuum \citep{francis93}.  To test this, we have created a
low-redshift geometric mean composite using only those quasars which
cover a rest wavelength of $5000$\,{\AA} ($z < 0.84$).  Since the
$1350${\AA} wavelength region we have used to measure continuum slopes
is not covered by the low-$z$ composite, we used instead the flux
density in the wavelength range $3020 - 3100${\AA} multiplied by a
factor of $0.86$, which is the ratio of the flux density of the
power-law fit to the flux density of the spectrum for the full-sample
geometric mean composite in that range.  We find a steeper index for
the low-redshift composite, $\alpha_\nu=-0.65$, than for the
full-sample composite, $\alpha_\nu=-0.44$, although the difference is
not as great as with the \citet{zheng97} composite.

Another apparently discrepant value is the result from
\citet{schneider01}, who find $\alpha_\nu=-0.93$ for a sample of very
high-redshift quasars.  Similar values for high-$z$ samples have been found
by \citet{fan01} and \citet{schneider91}.  The steep indices measured
for high-$z$ quasars may be due to the restricted wavelength range typically
used in fitting the continua, as suggested by \citet{schneider01}, and
not to a change in the underlying spectral index at high redshift.  At
high redshifts, only relatively short wavelength ranges redward
of \lya are avaliable in optical spectra, and these tend to be populated
by broad Fe\,{\sc ii} and Fe\,{\sc iii} complexes.  If, for example, the
regions of the median composite near $1350${\AA} and $1600${\AA} (just
redward of the \civ emission line) are taken as continuum (as
\citet{schneider01} did), we find a
power-law index of $\alpha_\nu=-0.93$.  This example demonstrates the
generic difficulty of measuring continuum indices without a very large
range of wavelength, or some estimate of the strength of the
contribution from blended emission lines.

The continuum slope changes abruptly near 5000\,{\AA} and becomes
steeper with an index of $\alpha_{\nu} = -2.45$, which is a good fit up
to the red end of the spectrum (8555\,{\AA}).  This change is also
evident in the FBQS composite, and has been noted in the spectra of
individual quasars \citep{wills85}.  An upturn in the spectral energy
distribution of quasars -- the so-called near-infrared inflection,
presumably caused by emission from hot dust -- has been seen starting
between $0.7$ and $1.5\mu$m \citep[e.g.][]{neugebauer87,elvis94}.  This
may be in part what we are seeing at wavelengths beyond $\approx
5000${\AA}, but it is unlikely that the sublimation temperature of dust
would be high enough for the emission to extend to wavelengths below
$6000${\AA} \citep{puget85,efstathiou95}.

Another possible contributor to the long wavelength steepening is
contamination from the host galaxies.  The $3\arcsec$ optical fiber
diameter subtends much if not all of the host galaxy image, even for
the lowest redshift quasars.  The best evidence for the contribution of
host galaxy light is the presence of stellar absorption lines in the
composite spectra.  The lines become stronger as the redshift, and
equivalently, luminosity, distributions of the quasar sample are
lowered.  This is seen by comparing the absorption line strengths of
the low-redshift median composite (\S\,\ref{emission}) with the
full-sample composite.  The strengths of the absorption lines in the
low-redshift median composite, assuming a typical elliptical galaxy
spectrum, imply a contribution to the composite quasar light from stars
of about $7 - 15\%$ at the locations of \caii\,$\lambda 3933$ and
Na\,{\sc i}\,$\lambda 5896$, and about $30\%$ at the locations of
Ca\,{\sc ii}\,$\lambda\lambda 8498, 8542$.  The trend of a greater
contribution from starlight with increasing wavelength is expected
because the least luminous quasars, in which the relative host galaxy
light is presumably most important, contribute the majority of spectra
to the composite at longer wavelengths.  This trend has also been seen
in the spectral light from the nuclei of individual low-redshift
Seyfert galaxies and other AGN \citep{terlevich90,seroteroos98}, which
suggest a significant contribution from starburst activity dominated by
red supergiants \citep{cidfernandes95}.  The mean absolute $r'$
magnitude of the quasars making up the low-$z$ composite is $M_{r'} =
-21.7$ (Fig.\,\ref{zdist}), which implies a host galaxy magnitude of
about $M_{r'} = -19.2$ (assuming a host contribution of $\sim 10\%$) --
a moderately luminous value in the SDSS filter system
\citep{blanton01}.  We conclude that both stellar light from the host
galaxies and a real change in the quasar continuum cause the steepening
of the spectral index beyond $5000${\AA}.

The detection of stellar Balmer absorption lines implies that young or
intermediate age stars make a substantial contribution to the light of
the host galaxies.  This is at odds with the conclusions based on
host-galaxy spectra \citep{nolan01}, and two-dimensional image
modeling \citep{mclure99,mclure00} that the hosts of quasars and radio
galaxies are ``normal'' giant ellipticals.  The discrepancy cannot
immediately be attributed to redshift differences, since the
\cite{mclure00} sample extends to $z\approx 1$, and we detect Balmer
absorption lines in the full-sample composite with a mean redshift of
$z=1.25$.  More likely, the difference is due to the fact that our
spectra include only the inner $3\arcsec$ of the galaxy light while the
spectra taken by \citet{nolan01} sample only off-nuclear ($5\arcsec$
from nucleus) light, and the image modeling includes the entire
profile of the galaxies.  This suggests that the stellar population
near the nuclei of quasar host galaxies -- near the quasars themselves --
is substantially younger than that of the host galaxies.

Velocity shifts in the BLR lines relative to the forbidden NLR lines --
taken to be at the systemic host-galaxy redshift -- are seen for most
quasars and are similar to the values we find for the composite BLR
lines relative to \oiii\,$\lambda 5007$
\citep[e.g.][]{tytler92,laor95,mcintosh99}.  The origin of the shifts
is not known, but explanations include gas inflows and outflows
\citep[e.g.][]{gaskell82, corbin90}, attenuation by dust
\citep{grandi77, heckman81}, relativistic effects \citep{netzer77,
corbin95, corbin97, mcintosh99}, and line emission from physically
different locations \citep[e.g.][]{espey89}.  The magnitudes of the
shifts seem to depend upon the ionization energies \citep{gaskell82,
wilkes86, espey89, tytler92, mcintosh99} in the sense that more
negative velocities (blueshifts) are seen for higher ionization lines.
We have confirmed this correlation using a large number of BLR lines
(\S\,\ref{shifts}).

It is often assumed that the NLR lines are at the systemic redshift of
the quasar, since the lines are thought to originate in a kpc-scale
region centered on the quasar, and the lines show good agreement (to
within $100$\,km\,s$^{-1}$) with the redshifts of host galaxies
determined by stellar absorption lines \citep{gaskell82,vrtilek85}, and
H\,{\sc i} 21\,cm observations \citep{hutchings87}.  However, for some
of the higher-ionization forbidden lines, such as \oiii\,$\lambda
5007$, Ne\,{\sc v}\,$\lambda3426$, Fe\,{\sc vii}\,$\lambda 6086$,
Fe\,{\sc x}\,$\lambda 6374$, and Fe\,{\sc xi}\,$\lambda 7892$,
seen in quasars and Seyfert galaxies, significant velocity shifts,
usually blueshifts, have been detected in the past
\citep[e.g.]{heckman81,mirabel84,penston84,whittle85,appenzeller91}.
The large number (17) of NLR lines we have been able to measure cover a
wide range in ionization potentials.  These lines are shifted with
respect to one another and the shifts are correlated with ionization
energy.  This appears to be a real effect, since we have been careful
to select only those lines which have well-defined non-blended peaks.
Another verification of the accuracy of the velocity measurements is
that lines originating from the same ion but at different rest
wavelengths almost always have consistent velocity offsets within the
measurement uncertainties (Table\,\ref{velshift} \&
Fig.\,\ref{velion}).

The NLR velocity shifts and their correlation with ionization potential
suggest that the same mechanism responsible for the shifts of the 
BLR lines also applies to the NLR lines, although the effect is weaker.
One possible explanation is that the BLR contains some lower density
forbidden line emitting gas, as first suggested by \citet{penston77}.
The correlation is strong, but the effect is subtle, so follow-up
work will likely have to involve both higher quality optical spectra and
observations in the near-IR in order to detect a sufficient sample
of narrow forbidden lines.

We have implicitly assumed that the velocity differences are
independent of other factors such as redshift and luminosity.  However,
\citet{mcintosh99} found that higher-$z$ quasars tend to have greater
velocity offsets relative to the \oiii line.  For quasars in our sample
with $z>0.84$, the \oiii emission line is redshifted out of the
spectra, which is why we used a cross-correlation technique to estimate
the center-of-mass redshifts.  If the true velocity offsets depend upon
redshift, the relation will be weakened by the cross-correlation
matching which finds the best match to a lower-redshift template, and
thus will tend to yield the lower-redshift emission line positions.  A
desirable future project is extending the wavelength
coverage to the near-infrared at higher redshift and to the ultraviolet
at lower redshift in order to simultaneously detect low and high
velocity lines.  Such a program of even a relatively modest sample size
would be highly beneficial to many quasar studies.

\section{Summary \label{summary}}
We have created median and geometric mean composite quasar spectra using a
sample of over 2200 quasars in the Sloan Digital Sky Survey.  The
resolution and signal to noise ratio exceed all previously published
UV/optical quasar composites.  Over 80 emission line features have been
detected and identified.  We have been able to measure velocity shifts
in a large number of both permitted and forbidden emission line peaks,
most of which have no such previous measurements.  Power-law fits to the
continua verify the results from most recent studies.  The composites
show that there is a lack of emission-free regions across most of the
UV/optical wavelength range, which makes fitting quasar continua difficult
unless a very wide wavelength range is available.

The SDSS is rapidly producing high-quality spectra of quasars which cover
a wide range of properties.  Composite spectra can therefore be made
from numerous sub-samples in order to search for dependencies of global
spectral characteristics on a variety of quasar parameters, such as
redshift, luminosity, and radio loudness -- a program which is
currently underway.  We are also using other techniques such as
principal component analysis to examine trends among the diversity of
quasar spectra.

The median composite is being used as a cross-correlation template for
spectra in the SDSS, and many other applications are imaginable.  The
median composite spectrum is likely to be of general interest, so it is
available as an electronic table (Table\,\ref{medianspec}).

\acknowledgments
The Sloan Digital Sky Survey (SDSS)\footnote{The SDSS Web site is
{\tt http://www.sdss.org/}.} is a joint project of The
University of Chicago, Fermilab, the Institute for Advanced Study, the
Japan Participation Group, The Johns Hopkins University, the
Max-Planck-Institute for Astronomy, New Mexico State University,
Princeton University, the United States Naval Observatory, and the
University of Washington. Apache Point Observatory, site of the SDSS
telescopes, is operated by the Astrophysical Research Consortium
(ARC).  Funding for the project has been provided by the Alfred P.
Sloan Foundation, the SDSS member institutions, the National
Aeronautics and Space Administration, the National Science Foundation,
the U.S. Department of Energy, Monbusho, and the Max Planck Society.
We thank Simon Morris for making an electronic version of the LBQS
composite quasar spectrum available to us, and Bev Wills for helpful
comments.  MAS acknowledges support of NSF grant AST-0071091.
DPS and GTR acknowledges support of NSF grant AST-990703.

\clearpage

\input{VandenBerk.tab1.tex}

\input{VandenBerk.tab2.tex}

\input{VandenBerk.tab3.tex}

\input{VandenBerk.tab4.tex}

\input{VandenBerk.tab5.tex}



\end{document}

%% file: VandenBerk.tab1.tex
\begin{deluxetable}{rrr}
\tablewidth{0pt}
\tablecaption{Median Composite Quasar Spectrum\tablenotemark{a}
\label{medianspec}}
\tablehead{
  \colhead{$\lambda$} &
  \colhead{$f_\lambda$} &
  \colhead{$f_\lambda$ Uncertainty} \\
  \colhead{({\AA})} &
  \colhead{(Arbitrary Units)} &
  \colhead{(Arbitrary Units)}
}
\tablecolumns{3}
\startdata
$800.5$ & $0.149$ & $0.074$ \\
$801.5$ & $0.000$ & $0.260$ \\
$802.5$ & $0.676$ & $0.227$ \\
$803.5$ & $0.000$ & $0.222$ \\
$804.5$ & $0.413$ & $0.159$ \\
$805.5$ & $0.338$ & $0.326$ \\
$806.5$ & $0.224$ & $0.159$ \\
$807.5$ & $0.122$ & $0.360$ \\
$808.5$ & $0.612$ & $0.346$ \\
$809.5$ & $0.752$ & $0.304$ \\
$810.5$ & $0.197$ & $0.257$ \\
$811.5$ & $0.187$ & $0.189$ \\
$812.5$ & $0.000$ & $0.126$ \\
$813.5$ & $0.000$ & $0.171$ \\
$814.5$ & $0.502$ & $0.181$ \\
\enddata
\tablecomments{The complete version of this table will appear in the
  electronic edition of the Journal.  The printed edition contains only
  a sample.  The full table can be found temporarily at
  {\tt ftp://sdss.fnal.gov/pub/danvb/qsocomposite/sdss\_qso\_median.tab1}}
\end{deluxetable}

%% file: VandenBerk.tab2.tex
\begin{deluxetable}{rrrrrrrll}
\tabletypesize{\scriptsize}
\tablewidth{0pt}
\tablecaption{Composite Quasar Emission Line Features \label{emlines}}
\tablehead{
  \colhead{$\lambda_{obs}$} &
  \colhead{$\lambda_{lo}$} &
  \colhead{$\lambda_{hi}$} &
  \colhead{Rel. Flux} &
  \colhead{$W$} &
  \colhead{Width} &
  \colhead{Skew} &
  \colhead{ID\tablenotemark{a}} &
  \colhead{$\lambda_{lab}$ ({\AA})} \\
  \colhead{({\AA})} &
  \colhead{({\AA})} &
  \colhead{({\AA})} &
  \colhead{100$\times F/F(\rm{Ly}\alpha)$} &
  \colhead{({\AA})} &
  \colhead{$\sigma_\lambda$({\AA})} &
  \colhead{} &
  \colhead{} &
  \colhead{or Multiplet\tablenotemark{b}}
}
\startdata
 940.93 $\pm$ 3.17 &  930 &  955 &   2.110 $\pm$ 0.379 &  2.95 $\pm$ 0.54 &  4.73 &  $0.59$ & Ly\,$\epsilon$: &  937.80 \\
                   &      &      &                     &                  &       &         & Ly\,$\delta$: &  949.74 \\
 985.46 $\pm$ 4.78 &  960 & 1003 &   5.195 $\pm$ 0.459 &  6.55 $\pm$ 0.58 &  8.95 & $-0.51$ &          CIII &  977.02 \\
                   &      &      &                     &                  &       &         &          NIII &  990.69 \\
1033.03 $\pm$ 1.27 & 1012 & 1055 &   9.615 $\pm$ 0.484 &  9.77 $\pm$ 0.49 &  7.76 & $-0.01$ &   Ly\,$\beta$ & 1025.72 \\
                   &      &      &                     &                  &       &         &           OVI & 1033.83 \\
1065.10 $\pm$ 5.09 & 1055 & 1077 &   0.816 $\pm$ 0.269 &  0.80 $\pm$ 0.27 &  4.17 &  $0.30$ &           ArI & 1066.66 \\
1117.26 $\pm$ 2.78 & 1100 & 1140 &   3.151 $\pm$ 0.289 &  3.66 $\pm$ 0.34 &  8.49 &  $0.02$ &        FeIII: &     UV1 \\
1175.35 $\pm$ 1.17 & 1170 & 1182 &   0.870 $\pm$ 0.148 &  0.83 $\pm$ 0.14 &  2.28 & $-0.01$ &         CIII* & 1175.70 \\
1216.25 $\pm$ 0.37 & 1160 & 1290 & 100.000 $\pm$ 0.753 & 92.91 $\pm$ 0.72 & 19.46 &  $0.40$ &  Ly\,$\alpha$ & 1215.67 \\
1239.85 $\pm$ 0.67 & 1230 & 1252 &   2.461 $\pm$ 0.189 &  1.11 $\pm$ 0.09 &  2.71 & $-0.21$ &            NV & 1240.14 \\
1265.22 $\pm$ 3.20 & 1257 & 1274 &   0.306 $\pm$ 0.081 &  0.21 $\pm$ 0.06 &  2.74 &  $0.25$ &          SiII & 1262.59 \\
1305.42 $\pm$ 0.71 & 1290 & 1318 &   1.992 $\pm$ 0.076 &  1.66 $\pm$ 0.06 &  5.42 & $-0.21$ &            OI & 1304.35 \\
                   &      &      &                     &                  &       &         &          SiII & 1306.82 \\
1336.60 $\pm$ 1.13 & 1325 & 1348 &   0.688 $\pm$ 0.059 &  0.59 $\pm$ 0.05 &  3.86 & $-0.02$ &           CII & 1335.30 \\
1398.33 $\pm$ 0.31 & 1360 & 1446 &   8.916 $\pm$ 0.097 &  8.13 $\pm$ 0.09 & 12.50 &  $0.06$ &          SiIV & 1396.76 \\
                   &      &      &                     &                  &       &         &          OIV] & 1402.06 \\
1546.15 $\pm$ 0.14 & 1494 & 1620 &  25.291 $\pm$ 0.106 & 23.78 $\pm$ 0.10 & 14.33 & $-0.04$ &           CIV & 1549.06 \\
1637.84 $\pm$ 0.83 & 1622 & 1648 &   0.521 $\pm$ 0.027 &  0.51 $\pm$ 0.03 &  4.43 & $-0.22$ &          HeII & 1640.42 \\
1664.74 $\pm$ 1.04 & 1648 & 1682 &   0.480 $\pm$ 0.028 &  0.50 $\pm$ 0.03 &  5.50 &  $0.11$ &         OIII] & 1663.48 \\
                   &      &      &                     &                  &       &         &          AlII & 1670.79 \\
                   &      &      &                     &                  &       &         &          FeII &    UV40 \\
1716.88 $\pm$ 2.83 & 1696 & 1736 &   0.258 $\pm$ 0.027 &  0.30 $\pm$ 0.03 &  7.36 &  $0.17$ &           NIV & 1718.55 \\
                   &      &      &                     &                  &       &         &          FeII &    UV37 \\
                   &      &      &                     &                  &       &         &          AlII & 1721.89 \\
1748.31 $\pm$ 0.75 & 1735 & 1765 &   0.382 $\pm$ 0.021 &  0.44 $\pm$ 0.03 &  5.12 &  $0.04$ &         NIII] & 1750.26 \\
1788.73 $\pm$ 0.98 & 1771 & 1802 &   0.229 $\pm$ 0.020 &  0.28 $\pm$ 0.02 &  6.06 & $-0.29$ &          FeII &   UV191 \\
1818.17 $\pm$ 2.07 & 1802 & 1831 &   0.130 $\pm$ 0.019 &  0.16 $\pm$ 0.02 &  5.72 & $-0.47$ &          SiII & 1816.98 \\
                   &      &      &                     &                  &       &         &       [NeIII] & 1814.73 \\
1856.76 $\pm$ 1.18 & 1840 & 1875 &   0.333 $\pm$ 0.021 &  0.40 $\pm$ 0.03 &  4.95 &  $0.01$ &         AlIII & 1857.40 \\
1892.64 $\pm$ 0.83 & 1884 & 1900 &   0.158 $\pm$ 0.015 &  0.16 $\pm$ 0.02 &  3.09 & $-0.10$ &        SiIII] & 1892.03 \\
                   &      &      &                     &                  &       &         &         FeIII &    UV34 \\
1905.97 $\pm$ 0.12 & 1830 & 1976 &  15.943 $\pm$ 0.041 & 21.19 $\pm$ 0.05 & 23.58 & $-0.27$ &         CIII] & 1908.73 \\
                   &      &      &                     &                  &       &         &         FeIII &     U34 \\
                   &      &      &                     &                  &       &         &         FeIII &    UV68 \\
                   &      &      &                     &                  &       &         &         FeIII &    UV61 \\
1991.83 $\pm$ 2.91 & 1976 & 2008 &   0.139 $\pm$ 0.014 &  0.20 $\pm$ 0.02 &  6.73 & $-0.03$ &         FeIII &    UV50 \\
2076.62 $\pm$ 0.78 & 2036 & 2124 &   1.580 $\pm$ 0.021 &  2.46 $\pm$ 0.03 & 16.99 &  $0.18$ &         FeIII &    UV48 \\
2175.62 $\pm$ 1.83 & 2153 & 2199 &   0.143 $\pm$ 0.013 &  0.25 $\pm$ 0.02 &  5.85 &  $0.46$ &          FeII &    UV79 \\
                   &      &      &                     &                  &       &         &          FeII &   UV370 \\
2222.29 $\pm$ 1.44 & 2202 & 2238 &   0.185 $\pm$ 0.011 &  0.33 $\pm$ 0.02 &  6.98 & $-0.11$ &          FeII &   UV118 \\
                   &      &      &                     &                  &       &         &          FeII &   UV376 \\
2324.58 $\pm$ 0.56 & 2257 & 2378 &   2.008 $\pm$ 0.020 &  3.56 $\pm$ 0.04 & 22.23 & $-0.29$ &          FeII &    many \\
2327.34 $\pm$ 0.72 & 2312 & 2338 &   0.183 $\pm$ 0.009 &  0.31 $\pm$ 0.02 &  4.95 & $-0.41$ &          CII] & 2326.44 \\
2423.46 $\pm$ 0.44 & 2402 & 2448 &   0.437 $\pm$ 0.012 &  0.77 $\pm$ 0.02 &  8.42 &  $0.25$ &        [NeIV] & 2423.83 \\
                   &      &      &                     &                  &       &         &         FeIII &    UV47 \\
2467.98 $\pm$ 1.59 & 2458 & 2482 &   0.092 $\pm$ 0.009 &  0.16 $\pm$ 0.02 &  4.54 &  $0.30$ &         [OII] & 2471.03 \\
                   &      &      &                     &                  &       &         &          FeII &   UV395 \\
2626.92 $\pm$ 0.99 & 2595 & 2654 &   0.398 $\pm$ 0.013 &  0.81 $\pm$ 0.03 &  9.93 &  $0.00$ &          FeII &     UV1 \\
2671.89 $\pm$ 1.78 & 2657 & 2684 &   0.067 $\pm$ 0.008 &  0.14 $\pm$ 0.02 &  5.10 &  $0.05$ &         AlII] & 2669.95 \\
                   &      &      &                     &                  &       &         &          OIII & 2672.04 \\
2800.26 $\pm$ 0.10 & 2686 & 2913 &  14.725 $\pm$ 0.030 & 32.28 $\pm$ 0.07 & 34.95 & $-0.06$ &          MgII & 2798.75 \\
2964.28 $\pm$ 0.79 & 2910 & 3021 &   2.017 $\pm$ 0.017 &  4.93 $\pm$ 0.04 & 22.92 & $-0.03$ &          FeII &    UV78 \\
3127.70 $\pm$ 1.07 & 3100 & 3153 &   0.326 $\pm$ 0.012 &  0.86 $\pm$ 0.03 &  9.38 & $-0.13$ &          OIII & 3133.70 \\
                   &      &      &                     &                  &       &         &          FeII &   Opt82 \\
3191.78 $\pm$ 0.99 & 3159 & 3224 &   0.445 $\pm$ 0.013 &  1.17 $\pm$ 0.03 & 12.77 & $-0.04$ &           HeI & 3188.67 \\
                   &      &      &                     &                  &       &         &          FeII &    Opt6 \\
                   &      &      &                     &                  &       &         &          FeII &    Opt7 \\
3261.40 $\pm$ 2.70 & 3248 & 3272 &   0.032 $\pm$ 0.008 &  0.09 $\pm$ 0.02 &  3.27 &  $0.06$ &           FeI &   Opt91 \\
                   &      &      &                     &                  &       &         &          FeII &    Opt1 \\
3281.74 $\pm$ 3.15 & 3272 & 3297 &   0.036 $\pm$ 0.008 &  0.10 $\pm$ 0.02 &  4.39 &  $0.56$ &          FeII &    Opt1 \\
3345.39 $\pm$ 0.75 & 3329 & 3356 &   0.118 $\pm$ 0.008 &  0.35 $\pm$ 0.02 &  5.50 & $-0.41$ &         [NeV] & 3346.82 \\
3425.66 $\pm$ 0.46 & 3394 & 3446 &   0.405 $\pm$ 0.012 &  1.22 $\pm$ 0.04 &  9.09 & $-0.62$ &         [NeV] & 3426.84 \\
3498.92 $\pm$ 1.60 & 3451 & 3537 &   0.432 $\pm$ 0.014 &  1.38 $\pm$ 0.05 & 16.79 & $-0.24$ &          FeII &    Opt4 \\
                   &      &      &                     &                  &       &         &          FeII &   Opt16 \\
3581.70 $\pm$ 4.48 & 3554 & 3613 &   0.100 $\pm$ 0.011 &  0.34 $\pm$ 0.04 &  7.98 &  $0.79$ &       [FeVII] & 3587.34 \\
                   &      &      &                     &                  &       &         &           HeI & 3588.30 \\
3729.66 $\pm$ 0.18 & 3714 & 3740 &   0.424 $\pm$ 0.009 &  1.56 $\pm$ 0.03 &  3.32 & $-0.24$ &         [OII] & 3728.48 \\
3758.46 $\pm$ 0.56 & 3748 & 3771 &   0.078 $\pm$ 0.007 &  0.29 $\pm$ 0.03 &  3.71 &  $0.12$ &       [FeVII] & 3759.99 \\
3785.47 $\pm$ 1.31 & 3775 & 3799 &   0.056 $\pm$ 0.006 &  0.22 $\pm$ 0.03 &  4.24 &  $0.13$ &         FeII: &   Opt15 \\
3817.41 $\pm$ 2.46 & 3800 & 3832 &   0.124 $\pm$ 0.007 &  0.51 $\pm$ 0.03 &  7.33 & $-0.10$ &         FeII: &   Opt14 \\
3869.77 $\pm$ 0.25 & 3850 & 3884 &   0.345 $\pm$ 0.008 &  1.38 $\pm$ 0.03 &  5.31 & $-0.50$ &       [NeIII] & 3869.85 \\
3891.03 $\pm$ 1.28 & 3882 & 3898 &   0.020 $\pm$ 0.005 &  0.08 $\pm$ 0.02 &  2.02 & $-0.27$ &           HeI & 3889.74 \\
                   &      &      &                     &                  &       &         &        H\,$8$ & 3890.15 \\
3968.43 $\pm$ 0.91 & 3950 & 3978 &   0.104 $\pm$ 0.007 &  0.45 $\pm$ 0.03 &  5.32 & $-0.62$ &       [NeIII] & 3968.58 \\
                   &      &      &                     &                  &       &         & H\,$\epsilon$ & 3971.20 \\
4070.71 $\pm$ 1.18 & 4061 & 4079 &   0.039 $\pm$ 0.005 &  0.18 $\pm$ 0.03 &  3.20 &  $0.01$ &         [FeV] & 4072.39 \\
                   &      &      &                     &                  &       &         &         [SII] & 4073.63 \\
4102.73 $\pm$ 0.66 & 4050 & 4152 &   1.066 $\pm$ 0.013 &  5.05 $\pm$ 0.06 & 18.62 &  $0.03$ &   H\,$\delta$ & 4102.89 \\
4140.50 $\pm$ 0.96 & 4135 & 4145 &   0.026 $\pm$ 0.004 &  0.13 $\pm$ 0.02 &  1.83 & $-0.38$ &          FeII &   Opt27 \\
                   &      &      &                     &                  &       &         &          FeII &   Opt28 \\
4187.55 $\pm$ 1.97 & 4157 & 4202 &   0.154 $\pm$ 0.009 &  0.76 $\pm$ 0.04 &  9.77 & $-0.40$ &          FeII &   Opt27 \\
                   &      &      &                     &                  &       &         &          FeII &   Opt28 \\
4239.85 $\pm$ 2.07 & 4227 & 4260 &   0.107 $\pm$ 0.008 &  0.53 $\pm$ 0.04 &  5.73 & $-0.05$ &        [FeII] &  Opt21F \\
4318.30 $\pm$ 0.78 & 4315 & 4328 &   0.038 $\pm$ 0.005 &  0.17 $\pm$ 0.02 &  2.31 &  $0.58$ &        [FeII] &  Opt21F \\
                   &      &      &                     &                  &       &         &          FeII &   Opt32 \\
4346.42 $\pm$ 0.38 & 4285 & 4412 &   2.616 $\pm$ 0.017 & 12.62 $\pm$ 0.08 & 20.32 &  $0.12$ &   H\,$\gamma$ & 4341.68 \\
4363.85 $\pm$ 0.68 & 4352 & 4372 &   0.110 $\pm$ 0.007 &  0.46 $\pm$ 0.03 &  3.10 & $-0.18$ &        [OIII] & 4364.44 \\
4478.22 $\pm$ 1.13 & 4469 & 4484 &   0.029 $\pm$ 0.006 &  0.14 $\pm$ 0.03 &  2.20 & $-0.69$ &          FeII &   Opt37 \\
                   &      &      &                     &                  &       &         &          HeI: & 4472.76 \\
4564.71 $\pm$ 1.56 & 4435 & 4762 &   3.757 $\pm$ 0.029 & 19.52 $\pm$ 0.15 & 61.69\tablenotemark{c} &  $0.23$ &          FeII &   Opt37 \\
                   &      &      &                     &                  &       &         &          FeII &   Opt38 \\
4686.66 $\pm$ 1.04 & 4668 & 4696 &   0.139 $\pm$ 0.009 &  0.72 $\pm$ 0.05 &  5.92 & $-0.57$ &          HeII & 4687.02 \\
4853.13 $\pm$ 0.41 & 4760 & 4980 &   8.649 $\pm$ 0.030 & 46.21 $\pm$ 0.16 & 40.44 &  $0.61$ &    H\,$\beta$ & 4862.68 \\
4930.75 $\pm$ 1.13 & 4920 & 4941 &   0.082 $\pm$ 0.007 &  0.40 $\pm$ 0.04 &  3.98 & $-0.02$ &          FeII &   Opt42 \\
4960.36 $\pm$ 0.22 & 4945 & 4972 &   0.686 $\pm$ 0.014 &  3.50 $\pm$ 0.07 &  3.85 & $-0.22$ &        [OIII] & 4960.30 \\
5008.22 $\pm$ 0.17 & 4982 & 5035 &   2.490 $\pm$ 0.031 & 13.23 $\pm$ 0.16 &  6.04 & $-0.22$ &        [OIII] & 5008.24 \\
5305.97 $\pm$ 1.99 & 5100 & 5477 &   3.522 $\pm$ 0.036 & 21.47 $\pm$ 0.22 & 74.83\tablenotemark{c} & $-0.25$ &          FeII &   Opt48 \\
                   &      &      &                     &                  &       &         &          FeII &   Opt49 \\
5160.81 $\pm$ 0.94 & 5149 & 5168 &   0.092 $\pm$ 0.008 &  0.52 $\pm$ 0.04 &  3.95 & $-0.42$ &       [FeVII] & 5160.33 \\
5178.68 $\pm$ 0.93 & 5170 & 5187 &   0.056 $\pm$ 0.007 &  0.32 $\pm$ 0.04 &  2.97 & $-0.23$ &        [FeVI] & 5177.48 \\
5201.06 $\pm$ 0.98 & 5187 & 5211 &   0.098 $\pm$ 0.008 &  0.56 $\pm$ 0.05 &  4.42 & $-0.37$ &          [NI] & 5200.53 \\
5277.92 $\pm$ 2.96 & 5273 & 5287 &   0.025 $\pm$ 0.007 &  0.14 $\pm$ 0.04 &  2.84 &  $0.38$ &       [FeVII] & 5277.85 \\
                   &      &      &                     &                  &       &         &          FeII &   Opt49 \\
5313.82 $\pm$ 1.85 & 5302 & 5328 &   0.118 $\pm$ 0.010 &  0.66 $\pm$ 0.06 &  5.34 &  $0.18$ &      [FeXIV]: & 5304.34 \\
                   &      &      &                     &                  &       &         &         [CaV] & 5310.59 \\
                   &      &      &                     &                  &       &         &          FeII &   Opt48 \\
                   &      &      &                     &                  &       &         &          FeII &   Opt49 \\
5545.16 $\pm$ 4.03 & 5490 & 5592 &   0.333 $\pm$ 0.021 &  2.12 $\pm$ 0.13 & 21.09 & $-0.04$ &      [ClIII]: & 5539.43 \\
                   &      &      &                     &                  &       &         &          FeII &   Opt55 \\
5723.74 $\pm$ 1.94 & 5704 & 5745 &   0.109 $\pm$ 0.015 &  0.70 $\pm$ 0.09 &  5.19 &  $0.72$ &       [FeVII] & 5722.30 \\
5877.41 $\pm$ 0.81 & 5805 & 5956 &   0.798 $\pm$ 0.029 &  4.94 $\pm$ 0.18 & 23.45 &  $0.26$ &           HeI & 5877.29 \\
6085.90 $\pm$ 2.00 & 6064 & 6107 &   0.113 $\pm$ 0.016 &  0.71 $\pm$ 0.10 &  3.78 & $-0.95$ &       [FeVII] & 6087.98 \\
6303.05 $\pm$ 0.53 & 6283 & 6326 &   0.179 $\pm$ 0.016 &  1.15 $\pm$ 0.11 &  3.14 & $-0.64$ &          [OI] & 6302.05 \\
6370.46 $\pm$ 2.67 & 6347 & 6400 &   0.217 $\pm$ 0.018 &  1.36 $\pm$ 0.11 & 10.18 & $-0.09$ &          [OI] & 6365.54 \\
                   &      &      &                     &                  &       &         &         [FeX] & 6376.30 \\
6551.06 $\pm$ 1.24 & 6544 & 6556 &   0.195 $\pm$ 0.029 &  0.43 $\pm$ 0.06 &  2.21 & $-0.20$ &         [NII] & 6549.85 \\
6564.93 $\pm$ 0.22 & 6400 & 6765 &  30.832 $\pm$ 0.098 & 194.52 $\pm$ 0.62 & 47.39 &  $0.35$ &  H\,$\alpha$ & 6564.61 \\
6585.64 $\pm$ 0.34 & 6577 & 6593 &   0.831 $\pm$ 0.034 &  2.02 $\pm$ 0.08 &  2.65 & $-0.02$ &         [NII] & 6585.28 \\
6718.85 $\pm$ 0.46 & 6708 & 6726 &   0.276 $\pm$ 0.014 &  1.65 $\pm$ 0.08 &  3.09 & $-0.27$ &         [SII] & 6718.29 \\
6733.72 $\pm$ 0.46 & 6726 & 6742 &   0.244 $\pm$ 0.013 &  1.49 $\pm$ 0.08 &  2.54 & $-0.11$ &         [SII] & 6732.67 \\
7065.67 $\pm$ 2.92 & 7034 & 7105 &   0.451 $\pm$ 0.026 &  3.06 $\pm$ 0.18 & 15.23 &  $0.05$ &           HeI & 7067.20 \\
7138.73 $\pm$ 1.12 & 7131 & 7148 &   0.082 $\pm$ 0.013 &  0.57 $\pm$ 0.09 &  3.13 & $-0.01$ &       [ArIII] & 7137.80 \\
7321.27 $\pm$ 3.55 & 7285 & 7360 &   0.359 $\pm$ 0.031 &  2.52 $\pm$ 0.22 & 14.26 &  $0.30$ &         [OII] & 7321.48 \\
7890.49 $\pm$ 3.33 & 7883 & 7897 &   0.096 $\pm$ 0.018 &  0.69 $\pm$ 0.13 &  2.97 &  $0.39$ &       [NiIII] & 7892.10 \\
                   &      &      &                     &                    &       &         &        [FeXI] & 7894.00 \\
\enddata
\tablenotetext{a}{Ions ending with a colon (:) are uncertain identifications.}
\tablenotetext{b}{Multiplet designations for Fe are from \citet{netzer83,
grandi81,phillips78}.}
\tablenotetext{c}{Broad feature composed mainly of Fe{\sc ii} multiplets.}
\end{deluxetable}

%% file: VandenBerk.tab3.tex
\begin{deluxetable}{rrrll}
\tabletypesize{\footnotesize}
\tablewidth{0pt}
\tablecaption{Composite Quasar Absorption Line Features \label{ablines}}
\tablehead{
  \colhead{$\lambda_{obs}$} &
  \colhead{$W$} &
  \colhead{Width} &
  \colhead{ID} &
  \colhead{$\lambda_{lab}$} \\
  \colhead{({\AA})} &
  \colhead{({\AA})} &
  \colhead{$\sigma_\lambda$({\AA})} &
  \colhead{} &
  \colhead{({\AA})}
}
\tablecolumns{5}
\startdata
\cutinhead{Median Composite Using All Spectra}
$3800.38 \pm 1.09$ & $0.35 \pm 0.03$ & $4.14$ & H\,$10$ & $3798.98$ \\
$3837.12 \pm 1.49$ & $0.46 \pm 0.03$ & $5.96$ &  H\,$9$ & $3836.47$ \\
$3934.96 \pm 0.55$ & $0.91 \pm 0.03$ & $7.11$ & Ca\,{\sc ii} & $3934.78$ \\
$8502.80 \pm 7.22$ & $1.11 \pm 0.61$ & $3.85$ & Ca\,{\sc ii} & $8500.36$ \\
$8544.17 \pm 1.89$ & $2.22 \pm 0.44$ & $3.87$ & Ca\,{\sc ii} & $8544.44$ \\
\cutinhead{Low-Redshift Median Composite ($z_{em} \leq 0.5$)}
$3737.82 \pm 1.03$ & $0.16 \pm 0.03$ & $0.97$ & H\,$13$:    & 3735.43 \\
$3749.45 \pm 1.13$ & $0.31 \pm 0.04$ & $2.96$ & H\,$12$:    & 3751.22 \\
$3774.09 \pm 1.27$ & $0.36 \pm 0.04$ & $3.56$ & H\,$11$     & 3771.70 \\
$3799.71 \pm 0.89$ & $0.84 \pm 0.05$ & $4.86$ & H\,$10$     & 3798.98 \\
$3837.77 \pm 1.16$ & $0.95 \pm 0.05$ & $5.69$ & H\,$9$      & 3836.47 \\
$3934.94 \pm 0.48$ & $1.64 \pm 0.06$ & $6.94$ & Ca\,{\sc ii} & 3934.78 \\
$3974.66 \pm 0.88$ & $0.36 \pm 0.04$ & $2.42$ & Ca\,{\sc ii}\tablenotemark{a} & 3969.59 \\
$5892.66 \pm 1.24$ & $0.44 \pm 0.05$ & $3.72$ & Na\,{\sc ii} & 5891.58 \\
                   &                 &        &             & 5897.56 \\
$8502.80 \pm 7.22$ & $1.11 \pm 0.61$ & $3.85$ & Ca\,{\sc ii} & 8500.36 \\
$8544.17 \pm 1.89$ & $2.22 \pm 0.44$ & $3.87$ & Ca\,{\sc ii} & 8544.44 \\
\enddata
\tablenotetext{a}{Contaminated by emssion from [Ne\,{\sc iii}]$\lambda 3967$
and H\,$\epsilon$.}
\end{deluxetable}

%% file: VandenBerk.tab4.tex
\begin{deluxetable}{lrrrrr}
\tabletypesize{\footnotesize}
\tablewidth{0pt}
\tablecaption{Emission line velocity shifts relative to
  [O\,{\sc iii}]$\lambda 5007$ \label{velshift}}
\tablehead{
  \colhead{Ion} &
  \colhead{$\lambda_{lab}$} &
  \colhead{$\lambda_{obs}$} &
  \colhead{$\Delta\lambda$} &
  \colhead{Velocity} &
  \colhead{Ionization} \\
  \colhead{} &
  \colhead{({\AA})} &
  \colhead{({\AA})} &
  \colhead{({\AA})} &
  \colhead{(km/s)} &
  \colhead{Energy (eV)}
}
\startdata
   Ly$\alpha$ & 1215.67 & 1216.25 & $ 0.58$ & \phn$143$ $\pm$ \phn91 &  13.60 \\
           NV & 1240.14 & 1239.85 & $-0.29$ & \phn$-70$ $\pm$ 162 &  97.89 \\
          CIV & 1549.06 & 1546.15 & $-2.91$ & $-563$ $\pm$ \phn27 &  64.49 \\
         HeII & 1640.42 & 1637.84 & $-2.58$ & $-471$ $\pm$ 151 &  54.42 \\
        NIII] & 1750.26 & 1748.31 & $-1.95$ & $-334$ $\pm$ 128 &  47.45 \\
        AlIII & 1857.40 & 1856.76 & $-0.64$ & $-103$ $\pm$ 190 &  18.83 \\
        CIII] & 1908.73 & 1907.30 & $-1.43$ & $-224$ $\pm$ \phn28 &  47.89 \\
         MgII & 2798.75 & 2800.26 & $ 1.51$ & \phn$161$ $\pm$ \phn10 &  15.04 \\
      $[$NeV] & 3346.82 & 3345.39 & $-1.43$ & $-128$ $\pm$ \phn67 & 126.22 \\
      $[$NeV] & 3426.84 & 3426.17 & $-0.67$ & \phn\phn$-58$ $\pm$ \phn38 & 126.22 \\
      $[$OII] & 3728.48 & 3729.66 & $ 1.18$ & \phn\phn$94$ $\pm$ \phn14 &  35.12 \\
    $[$FeVII] & 3759.99 & 3758.46 & $-1.53$ & $-122$ $\pm$ \phn44 & 125.00 \\
    $[$NeIII] & 3869.85 & 3869.77 & $-0.08$ & \phn\phn$-6$ $\pm$ \phn19 &  63.46 \\
    H$\delta$ & 4102.89 & 4102.73 & $-0.16$ & \phn$-11$ $\pm$ \phn48 &  13.60 \\
    H$\gamma$ & 4341.68 & 4342.02 & $ 0.34$ & \phn\phn$23$ $\pm$ \phn30 &  13.60 \\
     $[$OIII] & 4364.44 & 4364.15 & $-0.29$ & \phn$-19$ $\pm$ \phn48 &  54.94 \\
     H$\beta$ & 4862.68 & 4862.66 & $-0.02$ & \phn\phn$-1$ $\pm$ \phn14 &  13.60 \\
     $[$OIII] & 4960.30 & 4960.36 & $ 0.06$ & \phn\phn\phn$3$ $\pm$ \phn13 &  54.94 \\
     $[$OIII] & 5008.24 & 5008.22 & $-0.02$ & \phn\phn$-1$ $\pm$ \phn10 &  54.94 \\
    $[$FeVII] & 5160.33 & 5160.81 & $ 0.48$ & \phn\phn$27$ $\pm$ \phn54 & 125.00 \\
    $[$FeVII] & 5722.30 & 5722.27 & $-0.03$ & \phn\phn$-1$ $\pm$ 110 & 125.00 \\
          HeI & 5877.29 & 5876.75 & $-0.54$ & \phn$-27$ $\pm$ \phn56 &  24.59 \\
    $[$FeVII] & 6087.98 & 6086.90 & $-1.08$ & \phn$-53$ $\pm$ 151 & 125.00 \\
       $[$OI] & 6302.05 & 6303.05 & $ 1.00$ & \phn\phn$47$ $\pm$ \phn25 &  13.62 \\
      $[$NII] & 6549.85 & 6551.06 & $ 1.21$ & \phn\phn$55$ $\pm$ \phn56 &  29.60 \\
    H$\alpha$ & 6564.61 & 6565.22 & $ 0.61$ & \phn\phn$27$ $\pm$ \phn13 &  13.60 \\
      $[$NII] & 6585.28 & 6585.64 & $ 0.36$ & \phn\phn$16$ $\pm$ \phn15 &  29.60 \\
      $[$SII] & 6718.29 & 6718.85 & $ 0.56$ & \phn\phn$25$ $\pm$ \phn20 &  23.33 \\
      $[$SII] & 6732.67 & 6733.72 & $ 1.05$ & \phn\phn$46$ $\pm$ \phn20 &  23.33 \\
    $[$ArIII] & 7137.80 & 7138.73 & $ 0.93$ & \phn\phn$39$ $\pm$ \phn47 &  40.74 \\
\enddata
\end{deluxetable}

%% file: VandenBerk.tab5.tex
\begin{deluxetable}{lllrrl}
\tabletypesize{\footnotesize}
\tablewidth{0pt}
\tablecaption{Measurements of the optical power-law continuum index for
quasars. \label{indices}}
\tablehead{
  \colhead{$\alpha_\nu$} &
  \colhead{Sample} &
  \colhead{Measurement} &
  \colhead{Redshift} &
  \colhead{Median} &
  \colhead{Source} \\
  \colhead{} &
  \colhead{Selection} &
  \colhead{Method} &
  \colhead{Range} &
  \colhead{Redshift} &
  \colhead{}
}
\startdata
$-0.44$ & optical and radio  & composite spectrum             & $0.04 - 4.79$ & $1.25$ & (1) \\
$-0.93$ & optical            & average value from spectra     & $3.58 - 4.49$ & $3.74$ & (2) \\
$-0.46$ & radio              & composite spectrum             & $0.02 - 3.42$ & $0.80$ & (3) \\
$-0.43$ & radio              & composite spectrum (remeasure) & $0.02 - 3.42$ & $0.80$ & (3), (1) \\
$-0.39$ & radio              & photometric estimates          & $0.38 - 2.75$ & $1.22$ & (4) \\
$-0.33$ & optical            & average value from spectra     & $0.12 - 2.17$ & $1.11$ & (5) \\
$-0.99$ & optical and radio  & composite spectrum             & $0.33 - 3.67$ & $0.93$ & (6) \\
$-1.03$ & optical and radio  & composite spectrum (remeasure) & $0.33 - 3.67$ & $0.93$ & (6), (1) \\
$-0.46$ & optical            & photometric estimates          & $0.44 - 3.36$ & $2.00$ & (7) \\
$-0.32$ & optical            & composite spectrum             & NA\tablenotemark{a} & $1.3$ & (8) \\
$-0.36$ & optical            & composite spectrum (remeasure) & NA\tablenotemark{a} & $1.3$ & (8), (1) \\
$-0.67$ & optical            & composite spectrum             & $0.16 - 3.78$ & $1.51$ & (9) \\
$-0.70$ & radio              & composite spectrum             & NA\tablenotemark{a} & NA\tablenotemark{a} & (9) \\
\enddata
\tablenotetext{a}{The value was not given in the reference
nor derivable from the data.}
\tablerefs{
(1) This paper;
(2) \citet{schneider01};
(3) \citet{brotherton00};
(4) \citet{carballo99};
(5) \citet{natali98};
(6) \citet{zheng97};
(7) \citet{francis96};
(8) \citet{francis91};
(9) \citet{cristiani90}.
}
\end{deluxetable}